\documentclass[twocolumn,showpacs,preprintnumbers,amsmath,amssymb,floatfix]{revtex4}
\usepackage{graphicx}
\makeatother
\makeatletter

\begin{document}

\preprint{}

\title{Collective atomic recoil laser as a synchronization transition}

\author{J. Javaloyes}
\affiliation{Institut Mediterrani d'Estudis Avan\c{c}ats (IMEDEA),
Campus Universitat de les Illes Balears, E-07122 Palma de Mallorca, Spain.}
\author{M. Perrin}
\affiliation{Laboratoire de Physique des Lasers Atomes et
Mol\'ecules, F-59655 Villeneuve d'Ascq Cedex, France.\\
Max-Planck-Institut fur Physik Komplexer Systeme, Nothnitzer Str.
38, 01187 Dresden, Germany.}
\author{A Politi}
\affiliation{Istituto dei Sistemi Complessi, CNR,
via Madonna del Piano 10, I-50019 Sesto Fiorentino, Italy}

\date{\today}

\begin{abstract}
We consider here a model previously introduced to describe the collective 
behavior of an ensemble of cold atoms interacting with a coherent 
electromagnetic field. The atomic motion along the self-generated 
spatially-periodic force field can be interpreted as the rotation of a 
phase oscillator. This suggests a relationship with synchronization 
transitions occurring in globally coupled rotators. In fact, we show that
whenever the field dynamics can be adiabatically eliminated,
the model reduces to a self-consistent equation for the probability 
distribution of the atomic ``phases". In this limit, there exists a formal 
equivalence with the Kuramoto model, though with important differences
in the self-consistency conditions. Depending on the field-cavity
detuning, we show that the onset of synchronized behavior may occur through 
either a first- or second-order phase transition. Furthermore, we find
a secondary threshold, above which a periodic self-pulsing regime sets in,
that is immediately followed by the unlocking of the forward-field frequency.
At yet higher, but still experimentally meaningful, input intensities,
irregular, chaotic oscillations may eventually appear. 
Finally, we derive a simpler model, involving only five scalar variables,
which is able to reproduce the entire phenomenology exhibited by the
original model.
\end{abstract}
\pacs{42.50.Vk, 05.45.Xt, 05.65.+b., 42.65.Sf }
\maketitle

%
\section{Introduction}\label{Sec_Intro}

Much progress has been recently made in understanding the onset of collective
phenomena in cold atoms in the presence of a coherent electromagnetic field,
when atomic recoil cannot be neglected. In particular, the experimental
observation of collective atomic recoil laser (CARL) \cite{KCZC03}, accompanied
by the development of simple theoretical models \cite{PYN02} has revealed that
this is an appropriate physical environment for testing new and general ideas
on the behaviour of globally coupled oscillators. In fact, the position of
an atom moving along a line in a self-generated periodic potential can be
interpreted as a phase: this observation opens up the possibility to compare
CARL with other globally coupled systems \cite{KS75,BCM87,PRK96} and,
in particular, with the Kuramoto model (KM) \cite{KuraBook}. It is also
interesting to explore the formal analogy with neural networks, which are
currently the object of a strong research activity (see, e.g.
\cite{abbott,ZT04}) in the perspective of unraveling the underlying
information processing mechanisms. In fact, in one of the simplest, though
non-trivial modeling schemes, neurons can be assimilated to rotators, since
the action potential can be interpreted as a phase (this is, e.g., the case
of the so called ``leaky integrate and fire" neurons (LIF) \cite{abbott}). 
One of the
goals of this manuscript is precisely to investigate analogies and differences
between the collective phenomena that can arise in cold atoms and the
synchronization phenomena that occur in general models of globally coupled rotators.

Altogether, the idea of atomic recoil is often linked with optical
cooling \cite{CohenBook}, but several years ago it was suggested that the
recoil resulting from photon emission/absorption could induce macroscopic
phenomena \cite{carl} and possibly contribute to the transformation of
kinetic energy into coherent light emission, in analogy to what happens in the
free electron laser \cite{fel}.

However, for a long time, progress on the theoretical side was hindered
by the lack of a suitable model to describe the asymptotic stationary
behavior of an ensemble of atoms. Preliminary experiments conducted at
room temperature \cite{LBBT96,HBK96} were also partially inconclusive,
as pointed out in \cite{BGGV97}. As a consequence, it was unclear which
experimental conditions would be more appropriate for an experimental
observation and even whether collective phenomena could be seen at all.

With the introduction of the first model capable of accounting for stationary
phenomena \cite{PLP01}, it has been shown that at sufficiently low temperatures
(above the region where quantum effects become important) an atomic-polarization
grid can spontaneously arise, which triggers a coherent back-propagating
field \cite{JLP03}.
More recently, experimental evidence has been found \cite{KCZC03} and a corresponding
theory has been proposed \cite{JPLP04,RPFBCZ04,CSKZCRPB05}, of collective atomic
recoil lasing action in the presence of a very strong detuning, when the
atomic dynamics can be effectively described by a linear response theory.
Although a density grating arises in both cases, it is only in the latter setup
that it contributes to generating the back-propagating field, while in the
former one it just follows from the existence of two counterpropagating fields
and it is not the dominant mechanism providing self-amplification.

So far, the only collective behaviour that has been observed is the
spontaneous onset of a slightly-detuned backward-propagating constant field
through a second order transition. In this paper, we extend the theoretical
study, by first showing that the phase of the backward field can be eliminated
by referring to a frame that moves with the
instantaneous backward field frequency. This step allows uncovering an analogy
with the KM, but also some relevant differences. In both CARL and
KM, the force fields are self-consistently generated and depend on
the non-uniformity in the probability distribution; moreover, both of them
contain a mean-field sinusoidal term that eventually triggers ferromagnetic
ordering. However, at variance with the ``standard" KM, the slope
of the potential (i.e. the effective frequency of the oscillators) is
self-generated and there exists a preferred moving frame (the one we have
accurately selected) where the dynamics is autonomous. This makes it meaningful
to distinguish between locking (i.e.  perfect synchrony) and libration, depending
whether the potential exhibits local minima or not. 
The unavoidable presence of thermal noise introduces a further interesting effect,
namely, a mismatch between the average velocity of the atoms and that of the 
density grating. This indeed represents the starting point for establishing a 
connection with another transition (see below).

By exploring the behaviour for larger but still experimentally accessible
input intensities \cite{KCZC03}, we uncover an unexpectedly rich bifurcation
scenario, starting from the primary transition which appears to be either second
or first order, depending on the magnitude of the detuning between the injected
field and the nearest cavity mode.
More interesting is the secondary instability, which gives rise to amplitude
oscillations of both the backward and forward field and is then followed by
an unlocking transition, where the forward-field frequency too starts to be (red)
detuned with respect to the input field. On the one hand, this regime is reminiscent of
the periodic collective motion predicted in a model of charge density waves 
\cite{BCM87}, but it also resembles ``self-organized" quasiperiodicity, a phenomenon 
first found in a model of LIF neurons\cite{vanV}, revisited in \cite{MP} and proved in its
fully generality in \cite{PR07}. We show that all these features and the transition
to chaotic behaviour observed for yet higher input intensities are captured by a
simple model containing only five variables. Such a model is derived in the limit
of a strong cooperation parameter (see next section for its definition) and
weak dipolar coupling, when the probability density is well approximated by the 
first Fourier mode. However, its validity appears to extend to a wide parameter region. 
The simplified model proves useful also to unravel the nature of the primary transition 
in the presence of a nonzero cavity-field detuning.

The paper is organized as follows: In section \ref{sec_model}, we derive
the explicit form of the self-consistent washboard potential acting
on the particle probability density and discuss the analogies with the Kuramoto model.
In section \ref{sec_kura}, we provide a full characterization of the various
dynamical regimes appearing in this model. In section \ref{sec_simp}, we
derive a minimal model consisting of five ordinary differential equations,
that is able to reproduce the rich phenomenology of the full system. Finally,
in section \ref{sec_end}, we summarize the main results and comment about the
open problems.

\section{The model} \label{sec_model}

The original CARL model \cite{carl} considers only the dynamics of
the back-scattered field. Such approximation is valid in the
vicinity of the first threshold, but it fails at higher input
intensities. It becomes then necessary to consider the forward
mode dynamics as well, as first proposed in \cite{BRM97} and
further discussed in \cite{YN01,PYN02}. Besides, as analyzed in
\cite{JPLP04}, different models should be invoked, depending on
the physical mechanisms that are responsible for atomic
thermalization. For instance, if the process involves collisions
(with either an external buffer gas, or hard boundaries),
a Vlasov equation with a
BGK-type collisional operator \cite{BGK54} for the atomic density
in phase space \cite{BV96} is appropriate to model the
thermalization. This leads to a Vlasov-type equation for the
evolution of the density of probability. On the other hand, in the
context of cold atoms dynamics, thermalization can be achieved
{\it via} Doppler cooling \cite{cool}. Then, each cooling cycle
changes slightly the atomic momentum, and the appropriate
thermalization model, as shown in \cite{CohenBook}, may therefore
be a Fokker-Planck operator \cite{RiskenBook}, to describe the
interaction between the probability distribution of particles and
the molasses fields.

In this latter case, the model consists of a set of two equations
for the complex cavity fields $x_b$, $x_f$, coupled to a
Fokker-Planck equation for the single particle probability
distribution $Q(\theta,p)$ for the atomic position $\theta$ and
momentum $p$. Like in Ref.~\cite{JPLP04}, we limit ourselves to 
considering the strong friction limit, which is both
physically meaningful and allows to obtain some analytical
results.

Accordingly, under the simplifying hypothesis of a vanishingly small inertia,
the model reduces to a Fokker-Planck equation for the density 
$\rho(u,\overline t)$ of
atoms in the position $u$, accompanied by two equations for the complex
amplitudes $x_f$, $x_b$ of the forward and backward field, respectively
\begin{eqnarray}
\label{model1}
\partial_{\overline t} \rho &=& - \nu \partial_{u}\Im (x_f x_b^\star
{\rm e}^{2iu}) \rho + T \partial^2_{u} \rho, \nonumber \\
\frac{d{x}_f}{d\overline t} &=& -\kappa(1+ i \Delta)x_f + \kappa Y - i \kappa C x_b
\langle {\rm e}^{- 2iu} \rangle, \\
\frac{d{x}_b}{d\overline t} &=& -\kappa(1+i\Delta)x_b - i \kappa C x_f
\langle {\rm e}^{2iu} \rangle , \nonumber
\end{eqnarray}
where the adimensional parameters have the following meaning:
{\it i)} $C$ is the atom-field coupling constant;
{\it ii)} $\Delta$ is the suitably shifted (forward) field-cavity detuning;
{\it iii)} $\nu$ is the dipolar coupling;
{\it iv)} $T$ is the atomic temperature;
{\it v)} $Y$ is input field amplitude;
{\it vi)} $\kappa^{-1}$ is the photon lifetime within the ring
cavity, rescaled to the coherence time of the atomic transition.

We find it convenient to further rescale the variables, into 
$t=\kappa \overline t$, $x_f = YF$, $x_b= YB$, and $u = z/2$ 
(which implies $\rho(u,t) = 2 P(z,t)$). Accordingly, the model
reads,
\begin{eqnarray}
\label{model2}
\partial_{t} P &=& - \mu \partial_z \Im (F B^\star {\rm e}^{iz}) P +
\sigma \partial^2_z P, \nonumber \\
\dot F &=& -(1+ i \Delta) F + 1  -  i C B \mathcal{R}^\star,\\
\dot B &=& -(1+i\Delta) B  -  i C F \mathcal{R} ,\nonumber
\end{eqnarray}
where the dot denotes the derivative with respect to the new time variable $t$,
and we have explicitly introduced the order parameter,
\begin{equation}
\label{orderparam} \mathcal{R}(t) = \int_0^{2\pi} dz {\rm e}^{iz}
P(z,t).
\end{equation}
Moreover, we defined the two effective control parameters $\mu = 2
Y^2\nu/ \kappa$ and $\sigma = 4T/\kappa$. As a result, it turns
out that there are four relevant parameters that cannot be scaled
out, namely, the detuning $\Delta$, the so-called cooperation parameter $C$, the
input intensity $\mu$, and the temperature $\sigma$.\\

\subsection{A moving frame} \label{subsec_frame}

Next, we perform yet another change of variables to remove an irrelevant
variable (a phase) from the dynamics. We do that by referring to a moving
frame
\begin{equation}
\theta = z + \alpha(t),
\end{equation}
where $\alpha$ is to be defined. The new probability density writes,
\begin{equation}
Q(\theta,t) = P(z,t).
\end{equation}
Additionally, we introduce an amplitude-and-phase description for the two fields
\begin{eqnarray}
F &=& f {\rm e}^{i\phi}, \nonumber \\
B &=& b {\rm e}^{i\beta}. \nonumber
\end{eqnarray}
Notice that with these notations, a positive (negative) linear growth of the
phases is to be interpreted as a red (blue) shift in the field frequency.
Accordingly, the Fokker-Planck equation reads,
\begin{equation}
\partial_{t} Q = - \partial_\theta \left[ \dot \alpha +
\mu fb \sin (\phi-\beta-\alpha+\theta) \right] Q +
\sigma \partial^2_\theta Q.  \nonumber
\end{equation}
This equation suggests defining
\begin{equation}
\alpha \equiv \phi - \beta.
\end{equation}
By doing so, we obtain
\begin{equation}
\partial_{t} Q = - \partial_\theta \left[ \dot \phi - \dot \beta +
 \mu fb \sin \theta \right] Q + \sigma \partial^2_\theta Q.
\label{eq:FP-short}
\end{equation}
Moreover, the order parameter writes,
\begin{equation}
 \mathcal{R} = {\rm e}^{i(\beta-\phi)} R {\rm e}^{i\psi}
\end{equation}
where
\begin{equation}
R {\rm e}^{i\psi} \equiv R_c + i R_s = \int d\theta {\rm e}^{i\theta}  Q(\theta,t)
\end{equation}
As a result, the field equations write as
\begin{eqnarray}
\dot f &=& \cos \phi -f -CbR_s, \label{ampf}\\
\dot b &=& -b + Cf R_s, \label{ampb}\\
\dot \phi &=& -\frac{\sin \phi}{f} -\frac{CbR_c}{f} - \Delta, \label{phasef} \\
\dot \beta &=& -\frac{Cf R_c}{b} -\Delta. \label{phaseb}
\end{eqnarray}
We can now replace the expression for $\dot \phi$ and $\dot \beta$ in the
Fokker-Planck equation to finally obtain,
\begin{equation}
\label{fopl}
\partial_{t} Q = - \partial_\theta \left[ C \frac{f^2-b^2}{fb} R_c -\frac{\sin \phi}{f} +
 \mu fb \sin \theta \right] Q + \sigma \partial^2_\theta Q,
\end{equation}
from which we see that the variable $\beta$ does not play any role in
the dynamics, since it does not contribute to any of the force fields.
Accordingly, we conclude that the model is fully described by the three
equations (\ref{ampf},\ref{ampb},\ref{phasef}) plus the Fokker-Planck equation
(\ref{fopl}).
Upon interpreting $\theta$ as a phase, we can recognize atoms as a rotators
and the underlying dynamics as that of identical globally coupled
rotators in the presence of noise. The mutual interaction is mediated by the
two fields $F$ and $B$ which follow their own dynamics. The primary interest in
this setup was motivated by the possible existence of a regime where the
modulus of the order parameter (as well as the amplitude of the backward
field) is different from zero. In view of the above relationship with
rotator systems, the onset of this regime is akin to a 
synchronization transition.
However, it is also interesting to notice some analogies with the standard laser
threshold. In fact, in both cases the frequency of the backward field is
self-generated by the dynamics, but the corresponding phase does not contribute
to the dynamics itself. Accordingly, from a mathematical point of view, the
transition appears to be a degenerate Hopf bifurcation. Moreover, from the
above equations, it turns out that the reference frame moves with a velocity
equal to the frequency difference between the two fields. In dimensional
variables this means that the frame velocity is $2 (\dot \phi -\dot
\beta)/ k$, where $k$ is the wavenumber of the injected field.

\subsection{The physical parameter range} \label{subsec_param}

In order to keep contact with a possible experimental confirmation, we give here
the meaningful orders of magnitude of the four relevant parameters, making
reference to the experiment carried out in Ref.~\cite{KCZC03}. The cavity is
characterized  by a power transmission coefficient of the cavity mirrors
$\mathcal{T}=6.3\; 10^{-6}$ and a roundtrip length $\Lambda=8.5\; \text{cm}$.
Accordingly, the cavity linewidth is $\kappa=-c/\Lambda \ln(R) \sim 22$ kHz.
The atomic sample consists of $^{85} \text{Rb}$ atoms whose temperature
and density are $T_0=250 \mu$K and $n=3\; 10^{17} \text{m}^{-3}$,
respectively. The characteristic length of the atomic sample
is $L=10^{-3}\text{m}$. The optical parameters are given by the
coherence dephasing rate of the D1 transition, namely,
$2\gamma_{\perp} = \gamma_\parallel = 5.9$MHz.
The dipolar moment is $\mathcal{D}=1.5\;10^{-29}\text{cm}^{-1}$.
The detunings between the injected field and both the atomic transition
and the nearest cavity mode are $\Delta_a=1$ THz and $\Delta_c\sim0$,
respectively.
Therefore, the physical expressions and the relative orders of magnitude
of the parameters are
\begin{eqnarray}
C =&\frac{\alpha L}{\mathcal{T}} &\mathcal{O}(1 - 10^2), \nonumber \\
\Delta =& \Delta_c + C & \pm\mathcal{O} (0 - 10), \nonumber \\
\sigma =&\frac{k^2 k_B T_0}{m \kappa^2\gamma_\perp^2} &\mathcal{O}
(10^{-1} - 10^{1}),\nonumber \\
\mu =&\frac{Y^2 \bar\gamma \Omega}{2 \gamma K^2 \Delta_a}
 &\mathcal{O}(0 - 10), \nonumber
\end{eqnarray}
where $\alpha$ is the unsaturated absorption rate per unit length, 
$k_B$ is the Boltzman constant, and $m$ the atomic mass.

\section{Stationary states} \label{sec_kura}

Both in the perspective of determining the stationary solution and
to emphasize the analogies with the KM, we derive an {\it adiabatic
CARL model} (ACM) by setting the time-derivatives of the three field variables
equal to zero. In order to keep the notations as simple
as possible, initially we assume $\Delta=0$ (see Section V for the
qualitative changes induced by a nonzero detuning). From
Eq.~(\ref{ampb})
\begin{equation}
b= C R_s f.
\end{equation}
By then setting to zero the derivative in Eq.~(\ref{phasef}), we obtain
\begin{equation}
\sin \phi = -C^2 f R_s R_c,
\end{equation}
while, from Eq.~(\ref{ampf}),
\begin{equation}
\cos \phi = f(1 + C^2 R_s^2).
\end{equation}
By combining together these last two equations, one obtains
\begin{eqnarray}
f^{-2} = C^4 R_s^2R_c^2 + (1+C^2R_s^2)^2 .\nonumber
\end{eqnarray}
After replacing back into the equation for $Q(\theta,t)$, the
model reduces to
\begin{equation}
\partial_{t} Q = - \partial_\theta \left[ \omega(1 + \xi \sin \theta) \right] Q
 + \sigma \partial^2_\theta Q, \label{eq:selfcons_FP}
\end{equation}
where
\begin{eqnarray}
\omega &=&  \frac{R_c}{R_s} \label{eq:om}\\
\xi &=&  \frac{\mu CR_s^2}{R_c\left[(1+C^2R_s^2)^2+C^4R_s^2R_c^2\right]} \label{eq:xi}
\end{eqnarray}
Since the field dynamics is absent, the model resembles a typical
Fokker-Planck equation in a static potential, as studied extensively
by Risken \cite{RiskenBook}, except that the force field is here
determined self-consistently from some moments of the distribution $Q$.
The structure of the force field corresponds to that of the
so-called Adler, or washboard, potential \cite{adler}. The
parameters $\omega$ and $\xi$ respectively measure the tilt and
modulation amplitude of the potential. The tilt originates from
our choice of a moving frame: a stationary solution for the
probability density would indeed correspond to a moving grating in
the laboratory frame. The second parameter $\xi$ quantifies the
amplitude of the entraining force on the atomic could. For $\xi
\le 1$, the drifting velocity is simply modulated, but it does not
change sign. For $\xi \geq 1$, the washboard potential possesses a
local minimum and complete entrainment (synchronization) is
possible.

\subsection{The zero temperature limit} \label{subsec_zeroT}

In order to understand the underlying physics, it is convenient to start
from the zero-noise limit. A priori, the two most symmetric solutions are:
({\it i}) the perfectly synchronized state with all particles located in the
same position; ({\it ii}) the so-called splay state \cite{NW92}, characterized
by a constant flux of particles.

If all particles are located in $\theta$, then
$(R_c,R_s)=(\cos \theta,\sin \theta)$ and we can reduce the whole problem to the
ordinary differential equation
\begin{equation}
\dot \theta =  \frac{\cos \theta}{\sin \theta}  +
 \frac{\mu C \sin^2 \theta}{1+C^2(2+C^2)\sin^2 \theta}.
\end{equation}
The fixed point solution of the fully synchronized state is obtained by
determining the zero of the force field,
\begin{equation}
\cos \theta \bigg[1+C^2(2+C^2)\sin^2 \theta\bigg] =
 -\mu C \sin^3 \theta.
\end{equation}
By squaring it and introducing $X=\sin^2 \theta$, we obtain the equation
\begin{equation}
(1-X) \bigg[1+C^2(2+C^2)X\bigg]^2 = \mu^2 C^2 X^3.
\end{equation}
It is easy to verify that there exists a meaningful solution for any value of
the parameters $C$ and $\mu$. This means that, at zero temperature, any
arbitrarily small input field is able to trigger a backward field that is
sufficiently strong to entrain the atoms in the fully synchronized state.

On the other hand, the splay state is obtained by imposing that,
at zero temperature, the flux is constant, i.e. $\partial_\theta
[(\omega(1+\xi \sin \theta))Q]=\partial_t Q \equiv 0$. The
probability density then becomes proportional to the inverse of
the force field,
\begin{equation}
Q(\theta) = \frac{N}{\omega(1+\xi \sin \theta)},
\end{equation}
where $N$ is a normalization condition. From the definition of the
order parameter, Eq.~(\ref{orderparam}), we obtain the condition,
\begin{equation}
R_c+iR_s = \int d\theta \frac{{N\rm e}^{i\theta}}{\omega(1+\xi
\sin \theta)}.
\end{equation}
By solving the real part of this integral, one easily finds that $R_c=0$.
Accordingly, from Eq.~(\ref{eq:om}), it follows that $\omega=0$ which means
that there cannot be any tilt and, as a consequence, no splay state, because
the flux would be necessarily equal to zero.

\subsection{A comparison with the Kuramoto model} \label{subsec_spino}

At zero temperature, there exists a nontrivial collective state
for arbitrarily small input field. At finite temperature, as
already shown in Ref.~\cite{JPLP04}, there exists a threshold
value for the input intensity, below which, the noise washes out
any order.

The presence of such transition, as well as the underlying
structure of the model are reminiscent of what observed in KM. In
the absence of quenched disorder (i.e., assuming that all rotators
are identical), KM writes,
\begin{equation}
\partial_{t} Q = - \partial_\theta K R \sin (\theta-\psi) Q
 + \sigma \partial^2_\theta Q, \label{eq:kura1}
\end{equation}
where $K$ denotes the coupling constant, while the other notations are the same
as before. It is well known that the order parameter $R$ is larger than zero
only if the coupling constant is larger than some critical value which depends
on the noise strength.

From the comparison between ACM and KM, one can notice that the
sinusoidal force depends on the local phase in ACM, while it
depends on the phase difference in KM. This implies that the
dynamics of the KM is time independent in any moving frame (as
long as the velocity is constant). It is nevertheless convenient to
write the evolution equation in the frame which allows removing
the drift term (which is indeed absent in the KM). On the other
hand, the drift term cannot be removed from the ACM without
introducing an explicit time-dependence. A last difference
concerns the amplitude of the sinusoidal force which is
proportional to the order parameter in KM, while it is strongly
nonlinear, in the ACM, due to the coupling between order parameter
and field equations as it can be seen in Eqs.~(\ref{eq:om},\ref{eq:xi}).
It is now important to understand the implications of such
differences on the observed dynamics, especially in the presence
of stochastic processes, when phase-transitions are expected.

\subsection{The type of synchronization} \label{subsec_alpha}

\begin{figure}[htb]
\centering
\includegraphics[width=0.45\linewidth,height=0.45\linewidth, clip=true]
{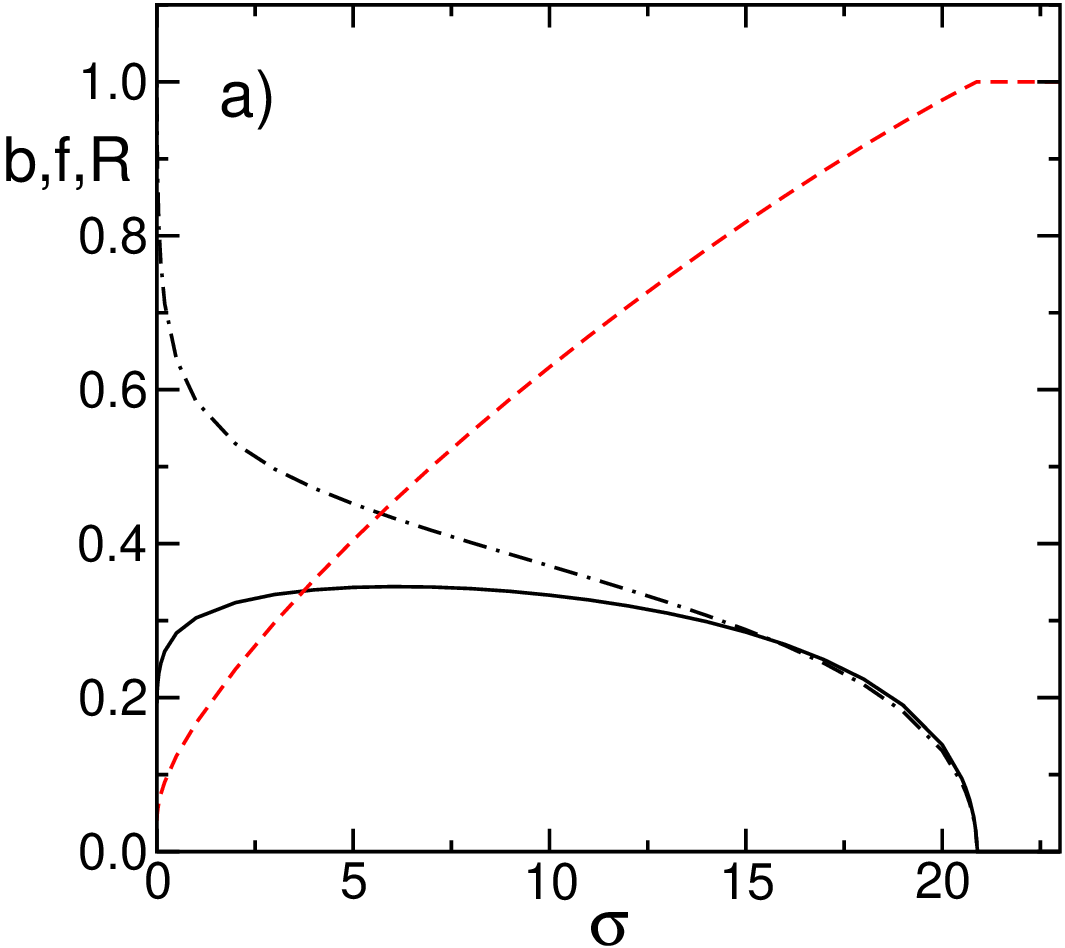}
\includegraphics[width=0.45\linewidth,height=0.45\linewidth, clip=true]
{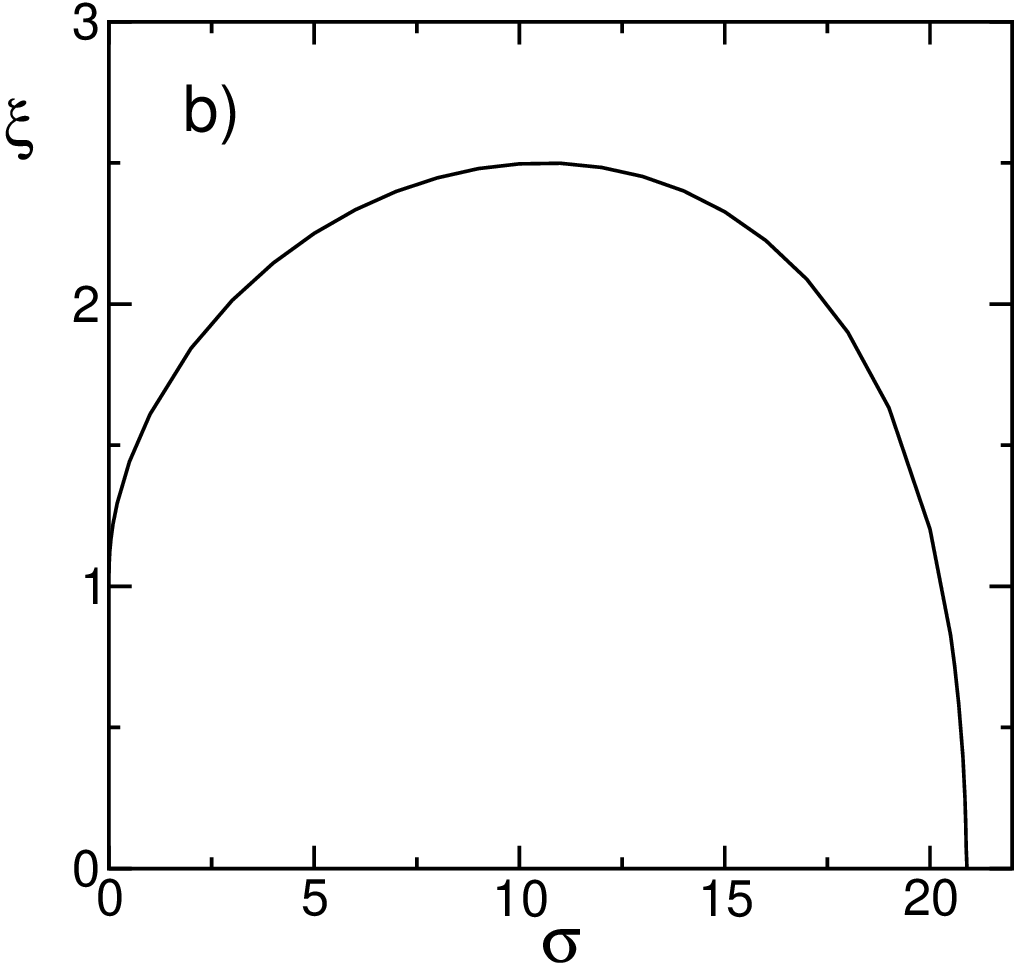} 
\caption{ (color online) Left panel: Bifurcation diagram
of the ACM model for $\mu=40$, $C=5$, $\Delta=0$. Solid, dashed,
and dotted-dashed lines correspond to $b$, $f$, and $R$ as a
function of $\sigma$. Right panel: the parameter $\xi$ of the
force field.} \label{fig:bifu_force}
\end{figure}

\begin{figure}[htb]
\centering
\includegraphics[width=0.9\linewidth, clip=true] {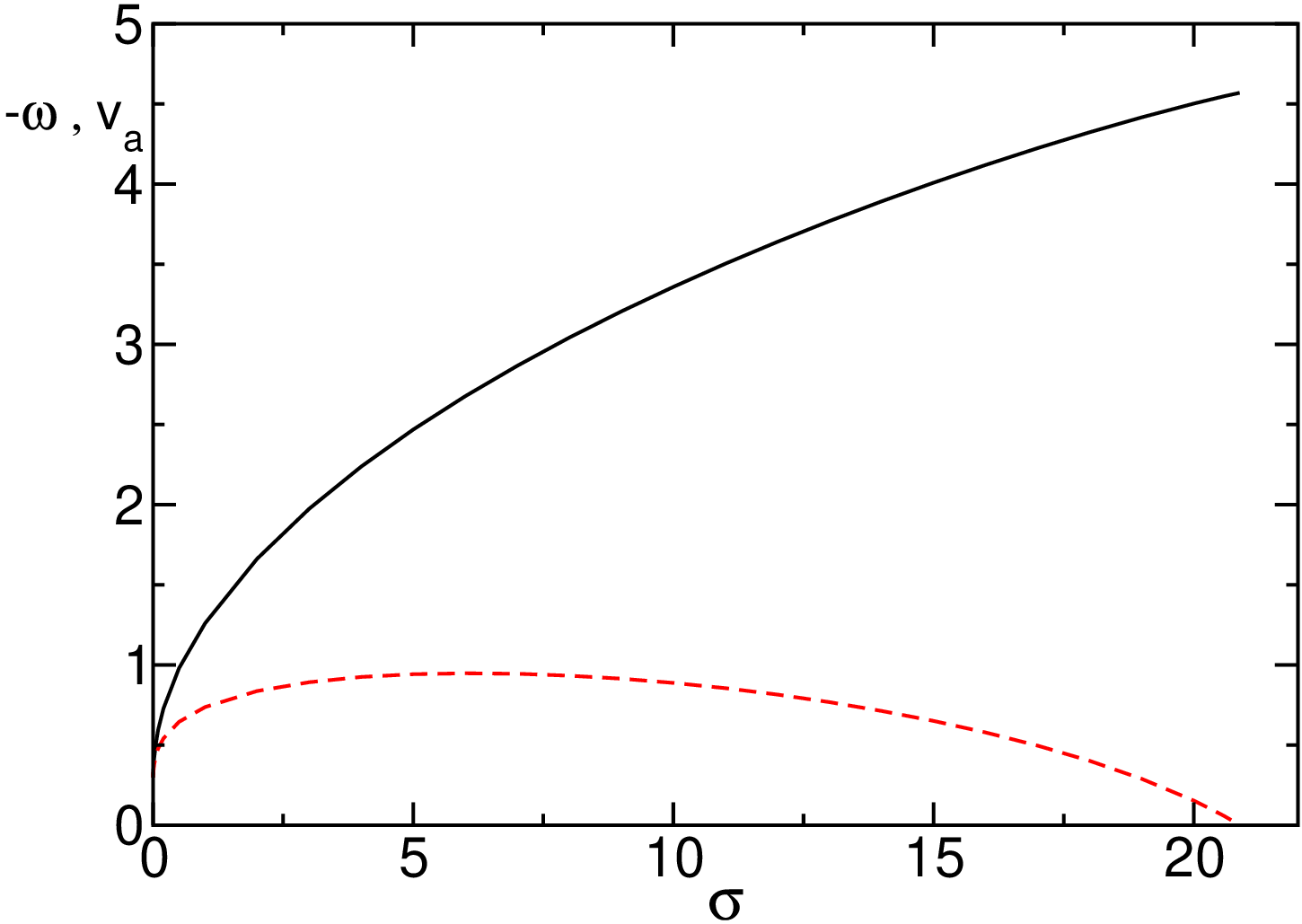}
\caption{ (color online) Velocity of the density grating ($-\omega$, solid line)
and average velocity of the atoms ($v_a$, dashed line) versus the scaled
temperature $\sigma$ and the same parameter values as in
Fig.~(\ref{fig:bifu_force}).} \label{fig:velo}
\end{figure}

In the vicinity of the primary transition, the backward field
intensity is arbitrarily small. Therefore, $\xi$ is also a small
quantity and the washboard potential cannot drag the atomic cloud.
All the variables being stationary, the flux is constant in the
vicinity of the transition, and the collective behavior is a
typical splay state. In order to understand how this regime
connects with the fully synchronous state observed in the
zero-noise limit, we solve numerically
Eqs.~(\ref{ampf},\ref{ampb},\ref{phasef},\ref{fopl}) for different
temperature values. As illustrated in Fig.~\ref{fig:bifu_force}a,
there is no backward field for large enough $\sigma$, while the
forward field intensity is constant and equal to 1 with our
normalizations. Upon decreasing $\sigma$ below the threshold
value, $f$ drops below 1 (dashed line), while $b$ increases from
zero (solid line) and, at small temperatures, decreases again in this
particular case. 
At the same time, the amplitude of the order parameter increases
monotonously from 0 to 1 (dot dashed line). Note that the highest
backward field does not correspond to the most coherent state
($R=1$), because of the nonlinear dependence of the potential
amplitude on the order parameter. At the same time, in
Fig.~\ref{fig:bifu_force}b, we see that for decreasing $\sigma$,
the relative amplitude $\xi$ of the modulation increases above 1
(meaning that the potential exhibits local minima) and eventually
decreases, though remaining larger than 1 at zero temperature,
when there is complete dragging.
On the other hand, we see in Fig.~\ref{fig:velo}, that the
velocity of the density grating, which is given by $-\omega$
(solid line), decreases monotonously with $\sigma$.
In the same figure, the dashed line corresponds to the average velocity $v$ of
the atoms, that is given by
\begin{equation}
v_a = -\omega + 2\pi\Phi \label{velocita}
\end{equation}
where $\Phi$ is the stationary flux of the Fokker-Planck equation
\begin{equation}
\Phi \equiv \omega Q(0) - \sigma \partial_\theta Q(0) .
\end{equation}
By comparing the two curves, we see that the density grating
velocity is everywhere larger than the atomic velocity except at
zero temperature, where the two coincide, sign of a complete
dragging. The difference is maximal in the vicinity of the primary
transition, where the atomic velocity is nearly zero (because the
backward field is negligible too), while the grating velocity is
maximal.

\begin{figure}[htp]
\centering
\includegraphics[width=0.95\linewidth,clip=true]{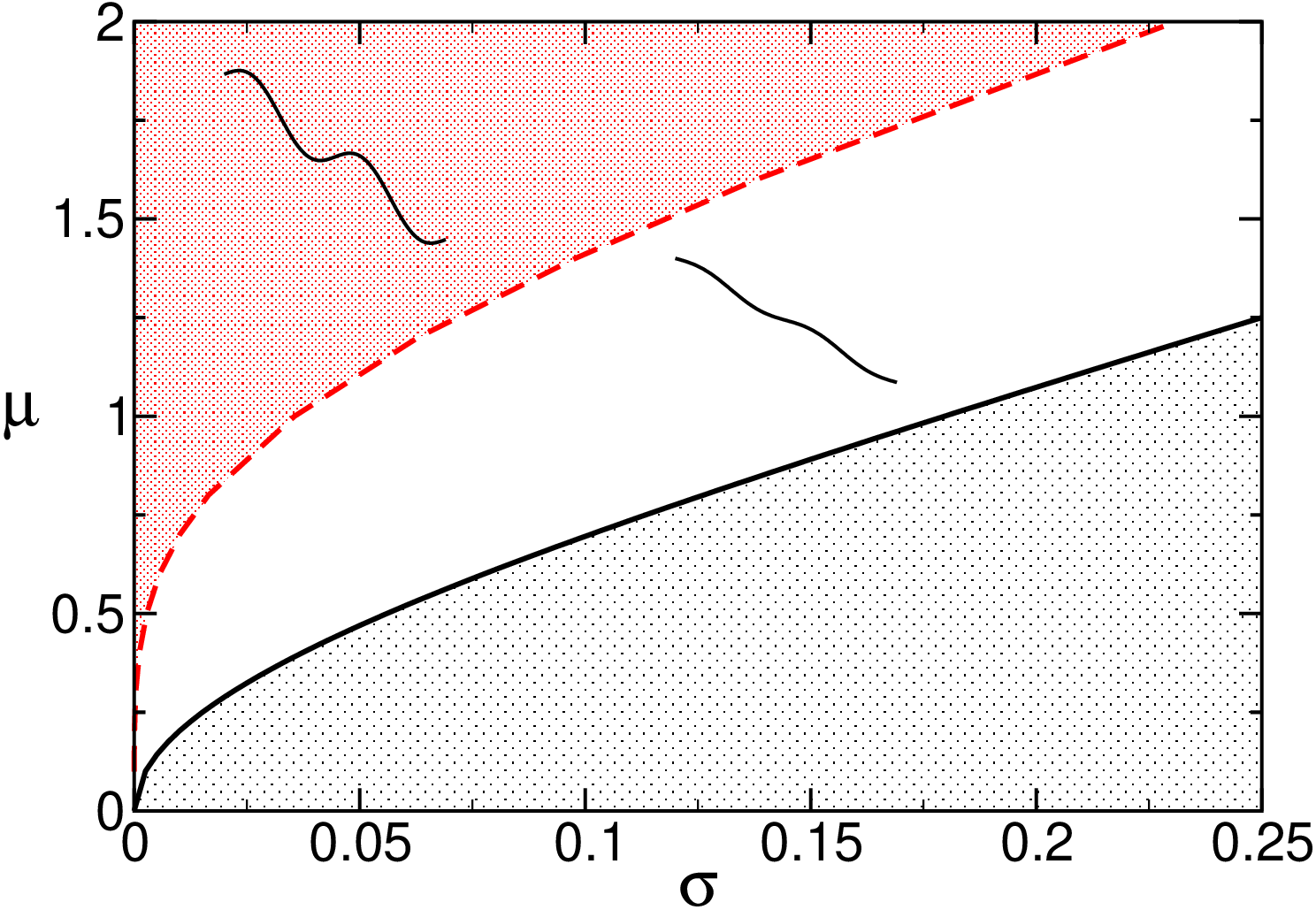}
\caption{(color online) Phase diagram separating locked/drifting from partially locked regimes for
$C=5$.}
\label{fig:phasediag}
\end{figure}

In order to check the generality of this scenario, we present in
Fig.~\ref{fig:phasediag} a phase diagram for different values
of both $\sigma$ and the input intensity $\mu$. There we see that
above the transition line,
there are two broad areas that extend down to $\sigma=0$. In the
first one (color-shaded, for larger values of $\mu$), the
potential has minima, and there is no fixed point. In the second
one, (in white, between the transition line and the dashed line),
the potential has no minima. The dashed line separating these two
regions does not represent a true transition line, but marks a
quantitative difference. above (first region), the flux is
triggered by the noise, since the barrier to the right of a
minimum is lower than that on the left; below (second region), the
flux is intrinsically deterministic.

\section{Numerical analysis} \label{subsec_Numana}

\begin{figure}[htp]
\centering
\includegraphics[width=0.9\linewidth, clip=true]{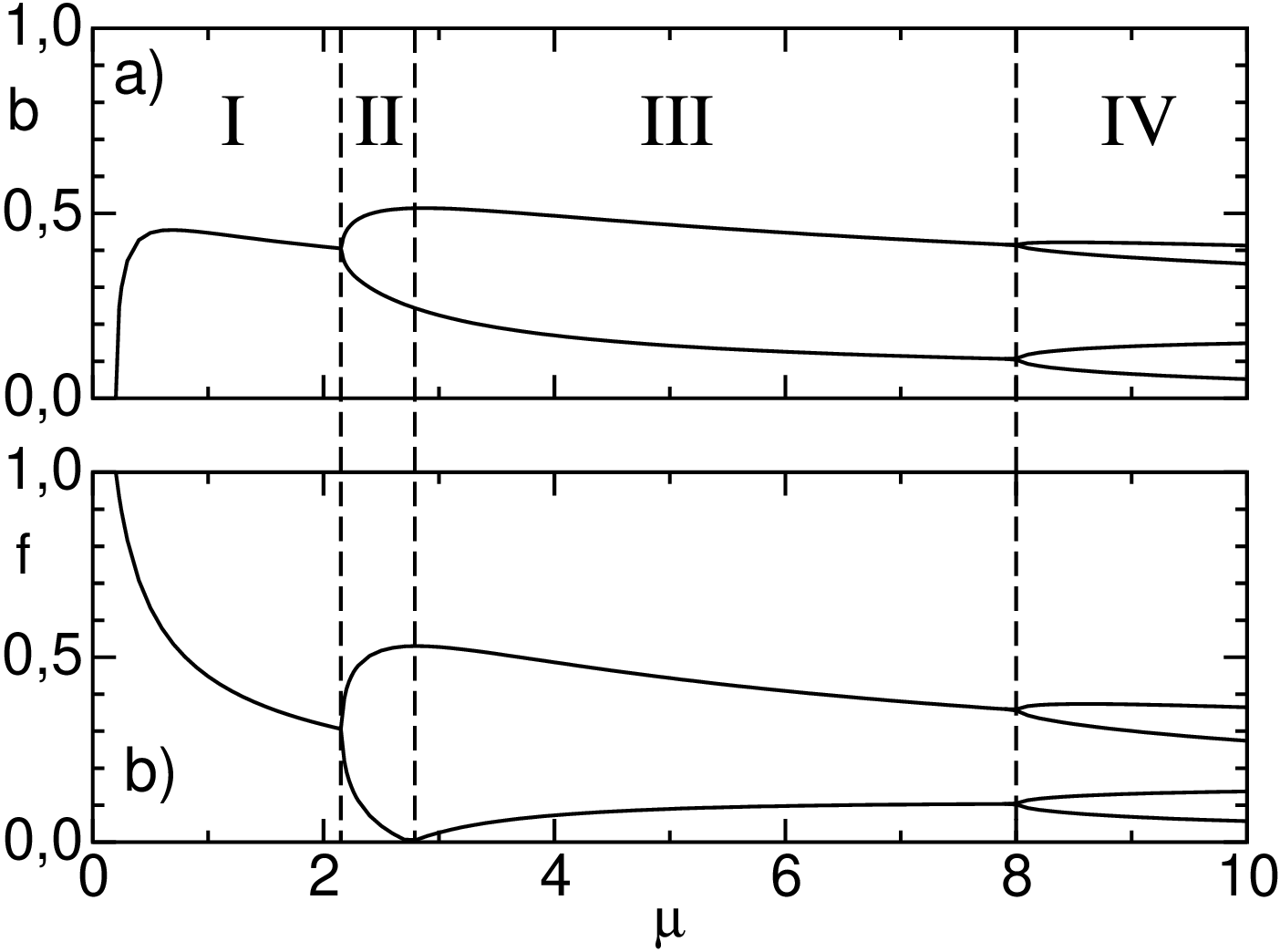}
\caption{
Bifurcation diagram for the backward (panel a) and forward (panel b) field
amplitudes as a function of $\mu$. The other parameters are $C=20$, $\sigma=1$,
and $\Delta=0$. The meaning of the various regions is discussed in the
text.}
\label{fig:2ndary_bifudiag}
\end{figure}

So far we have discussed the stationary state that arises from the solution
of the ACM. We have shown the existence of two phases: {\it i)} a trivial one
characterized by a zero order parameter and an independent evolution of the
single atoms; {\it ii)} a collective state characterized by two different
velocities for the single atoms and the density grating.
This scenario is superficially reminiscent of that one found in the KM,
but the properties of the collective motion are slightly more subtle.
In the frame where the dynamics is stationary, there is still a
non-zero flux induced by a finite tilting, that cannot be removed.
However, there are further differences. At variance with
the KM, here we show the existence of more complicated dynamical
regimes that appear when the amplitude of the injected field is
further increased beyond the primary transition. In
Fig.~\ref{fig:2ndary_bifudiag}, we show the typical sequence of
states that are detected upon increasing the input field $\mu$.

\begin{figure}[htp]
\centering
\includegraphics[width=0.9\linewidth, clip=true]{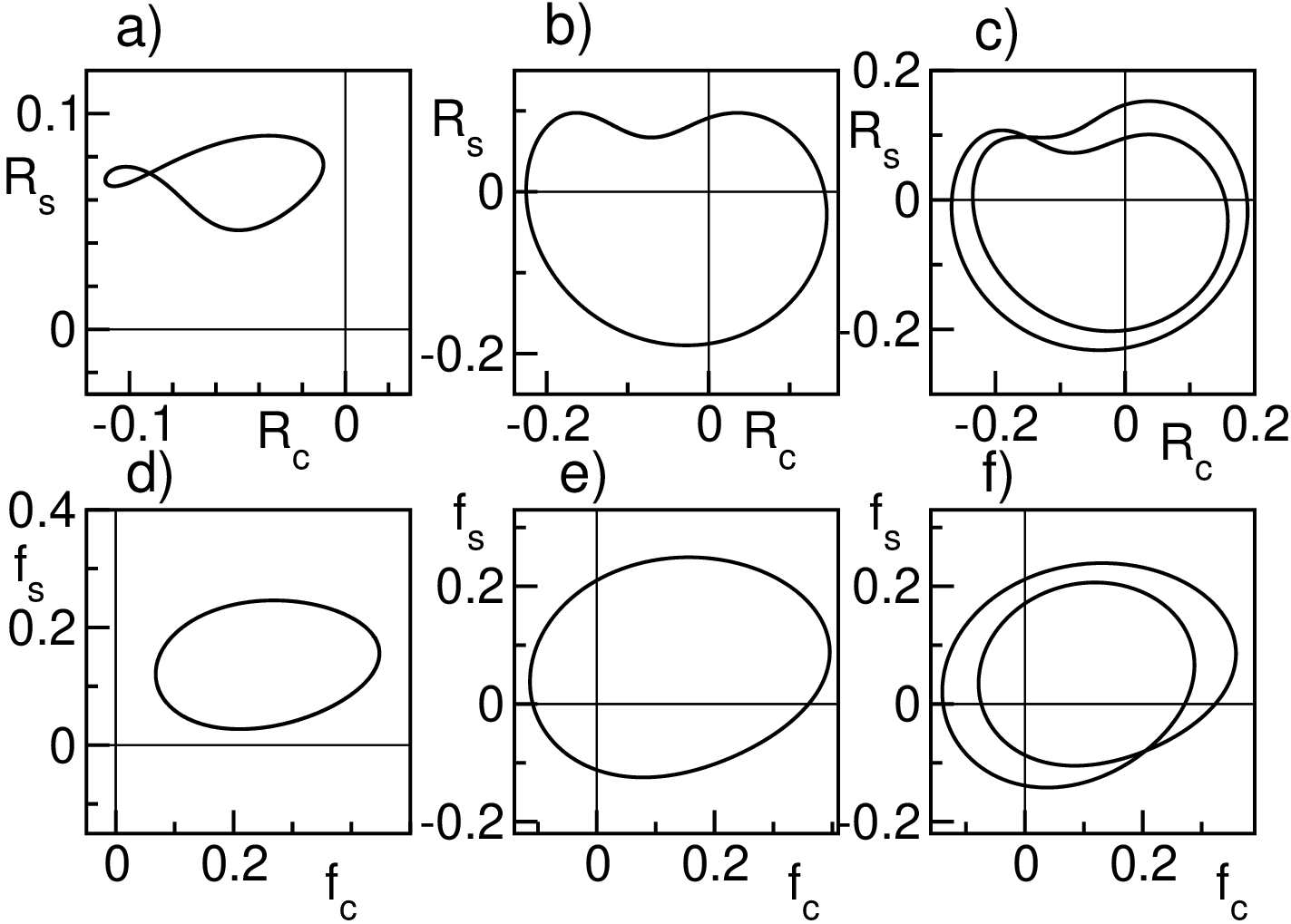}
\caption{ Phase space portraits of the typical behaviour observed
in region $II$ (panels a,d, $\mu=2.3$), III (panels b,e, $\mu=6$)
and region $IV$ (panels c,f $\mu=9$). Parameters are those of
Fig.~(\ref{fig:2ndary_bifudiag}).} \label{fig:cycle}
\end{figure}

For each value of $\mu$, the extrema of the field are plotted.
Inside region $I$, there is only one point,
meaning that the collective state is stationary in the moving
frame. A secondary Hopf bifurcation separates region $I$ from region $II$,
where the two field amplitudes start oscillating. In region $II$,
the average frequency of the forward field still remains locked to that of the
input field. This can be appreciated in Fig.~\ref{fig:cycle}a,d
where we plot the real and imaginary components of the order parameter
($R_c$, $R_s$) and of the forward field ($f_c$, $f_s$) for $\mu=2.3$.
In fact, neither $R$, nor $F$ exhibit an overall rotation since, in both
representations, the limit cycle does not encircle the origin.
Upon further increasing $\mu$, an unlocking
occurs: in region $III$, the periodic oscillation amplitudes are
accompanied by a rotation, as it can be seen in
Fig.~\ref{fig:cycle}b,e, where both limit cycles projections now
encircle the origin for $\mu=6$ (see Sec.~\ref{Sec:unlocking} for details).
Finally, a period doubling
bifurcation signals the appearance of yet more complicated
dynamical states and the possible onset of a chaotic dynamics.
This occurs in region $IV$ and can be appreciated by looking at
Fig.~\ref{fig:cycle}c,f, where the phase state projections are
plotted for $\mu=9$.

\subsection{The secondary transition} \label{subsec_second}

Besides solving directly the Fokker-Planck equation, we have
determined the locus of the secondary Hopf bifurcation, by first
determining the steady state in terms of a continued fraction
expansion as described in \cite{JPLP04}. The equilibrium
probability distribution has been evaluated analytically by using
its integral form (see e.g. \cite{RiskenBook} for further details);
the stability analysis has been thereby carried out by introducing
infinitesimal perturbations for the fields and for the probability
distribution, by referring to an equispaced mesh containing
$N\sim 256$ points. Finally, we have determined the eigenvalues of
a sparse Jacobian matrix of size $(N+3)^2$ by means of the QR
decomposition \cite{meshchach} while the integral form of the
equilibrium distribution has been evaluated by using the
Clenshaw-Curtis quadrature integration scheme \cite{Flam}.
\begin{figure}[htp]
\centering
\includegraphics[width=0.95\linewidth, clip=true]{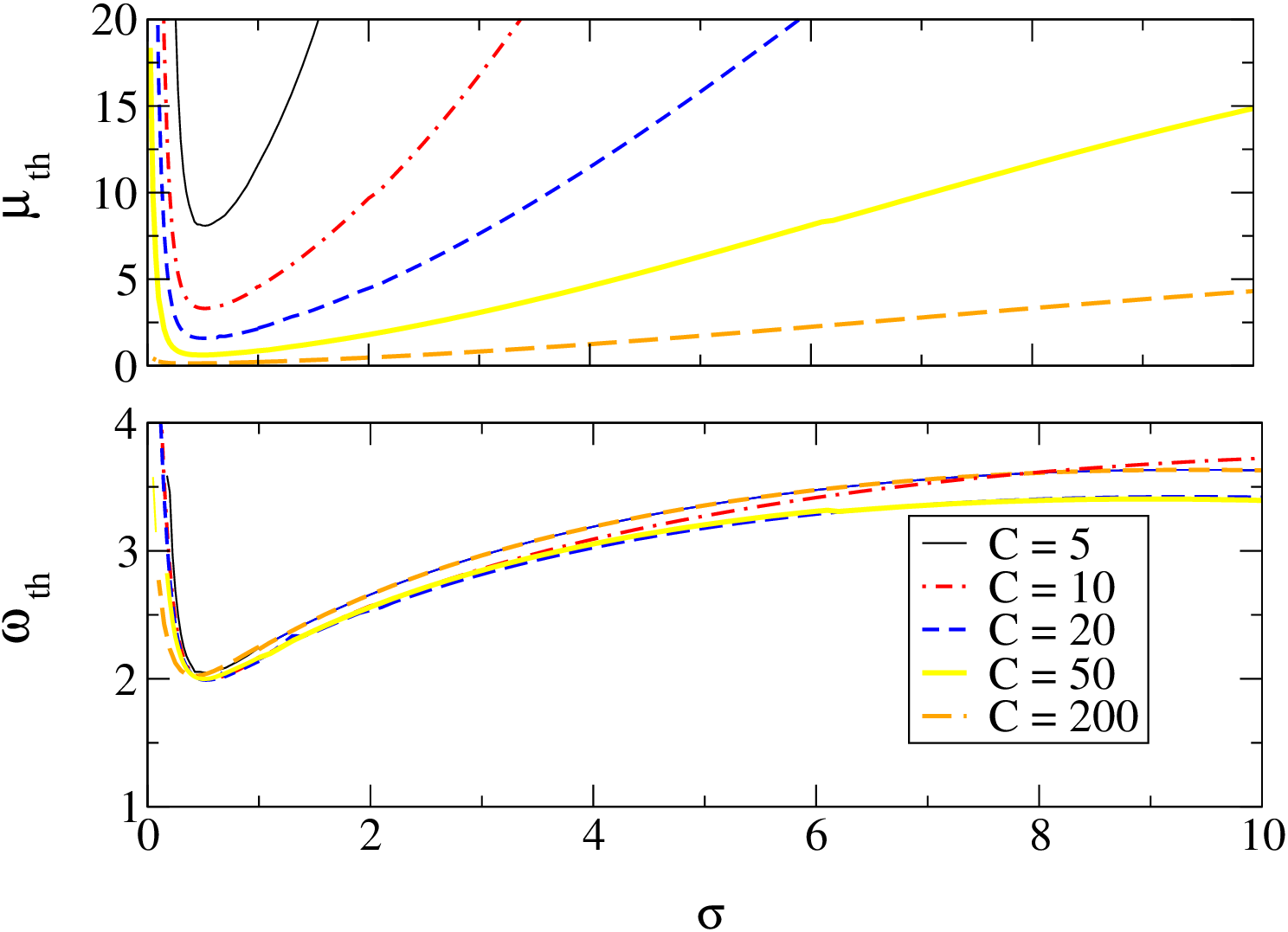}
\caption{ (color online) Locus of the secondary Hopf bifurcation as a function
$\mu$ and $\sigma$, for different values of $C$. As $C$ decreases,
the threshold for self-oscillation goes toward infinite values of
$\mu$. } \label{fig:2ndary_hopf}
\end{figure}

The main results are summarized in Fig.~\ref{fig:2ndary_hopf}.
There, one can see that the secondary bifurcation is inhibited at small
temperatures, where both the threshold power $\mu_h$ and the Hopf frequency $\omega_h$
diverge to infinity. This inhibition also takes place for small values
of the cooperation parameter $C$.

Furthermore, it is worth noticing the limiting behavior of the
secondary threshold as $C$ is increased: All bifurcation curves
tend to accumulate on an asymptote. Besides, the frequency of the
secondary bifurcation is almost independent of $C$. This suggests
that in the large-$C$ limit, the parameter $C$ can be scaled out
of the problem. In fact, in the next section we show that it is
convenient to introduce the smallness parameter $1/C$ and thereby
suitably expand the dynamical equations.

\begin{figure}[htp]
\centering
\includegraphics[width=0.9\linewidth, clip=true]{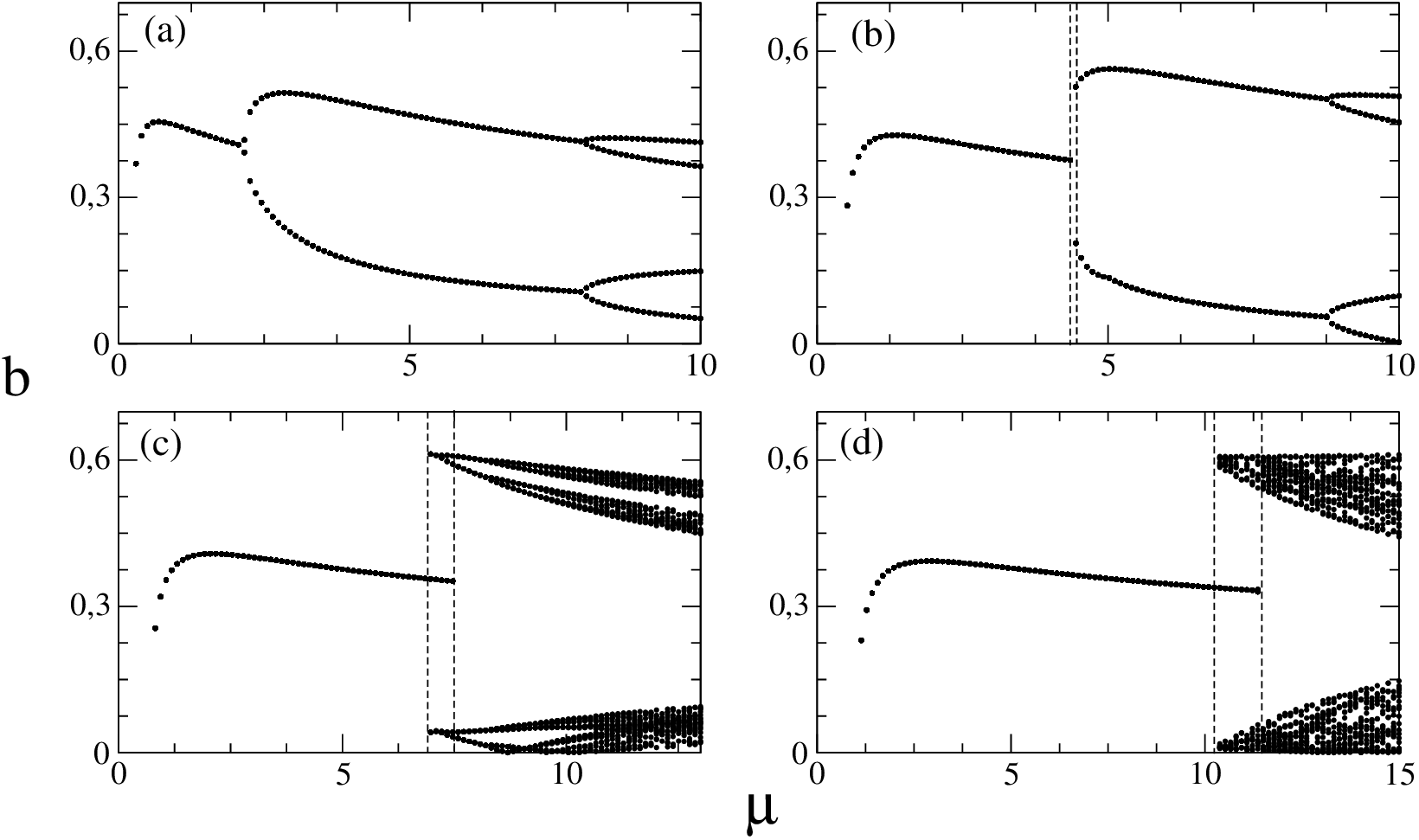}
\caption{ Bifurcation diagram for the backward
field amplitude $b$ as a function $\mu$. the fixed parameters are
$C=20$ and $\Delta=0$. Panels (a),(b),(c), and (d) correspond 
to $\sigma=1,2,3$ and $4$, respectively. As $\sigma$ increases, a region of
bistability indicated by the vertical dashed lines, becomes more and more visible thus indicating a
subcritical secondary Hopf bifurcation.}
\label{fig:2ndary_bifu4diag}
\end{figure}

Depending on the temperature value $\sigma$, we have found that the
secondary bifurcation can be either super- or sub-critical, as it can
be appreciated in Fig.~\ref{fig:2ndary_bifu4diag} where a narrow but
increasing region of bistability can be identified for the two larger $\sigma$
values. The diagrams have been obtained by sweeping the control parameter $\mu$
both in the increasing and decreasing direction, while keeping constant
the parameters $C=20$ and $\Delta=0$.

Finally, in order to clarify whether the field dynamics 
is a necessary ingredient
for the richness of the observed phenomenology, we artificially reintroduced the
photon lifetime into the field equations by multiplying their time derivatives
by $\kappa$. The field dynamics can thereby be adiabatically eliminated in the
limit $\kappa \rightarrow \infty$. By running extensive numerical simulations,
we have found that when $\kappa$ increases, the Hopf bifurcation occurs for
diverging values of both $\mu$ and $\sigma$. This provides an indirect indication
that field dynamics is a necessary ingredient for observing the essential
features of the bifurcation scenario.

\subsection{The unlocking transition} \label{Sec:unlocking}
\begin{figure}[htp]
\centering
\includegraphics[width=0.85\linewidth, clip=true]{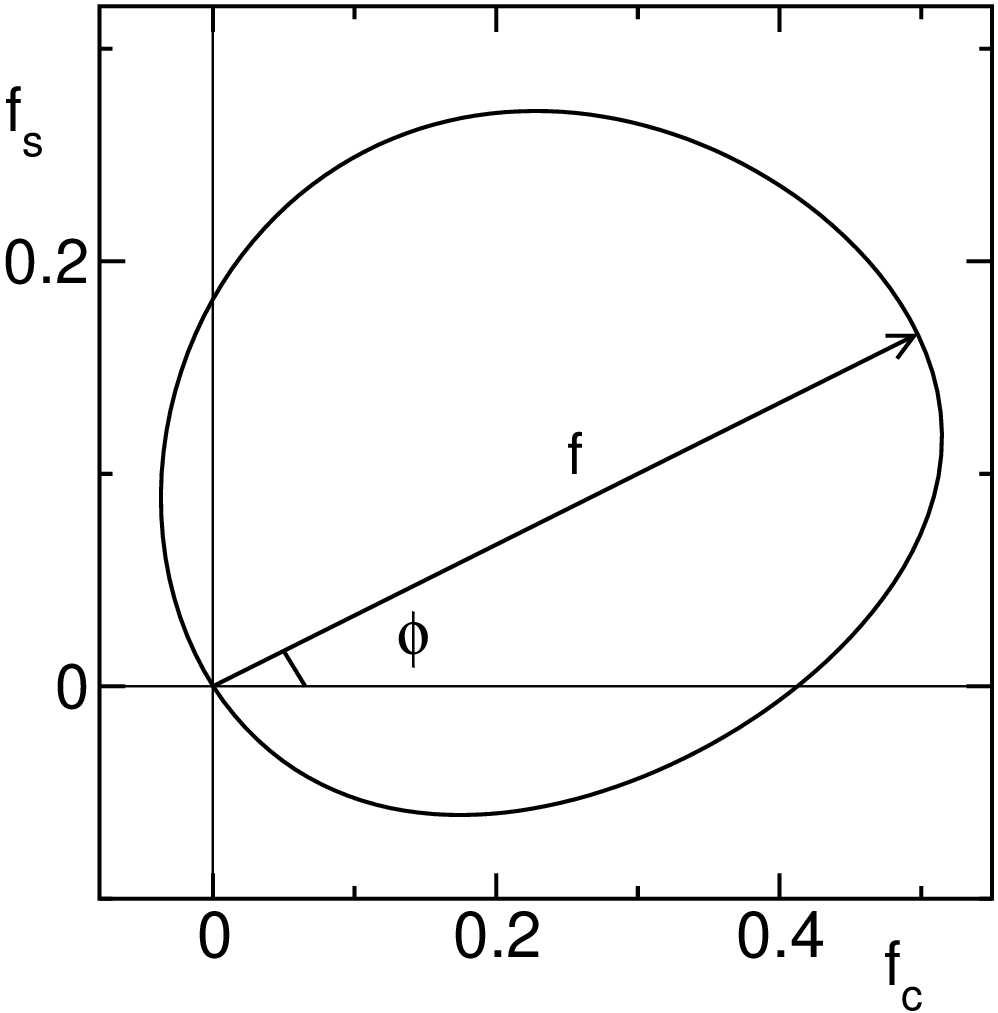}
\caption{ A projection of $F$ in the complex plane, for the
critical value $\mu_c=2.752$, when the forward field unlocks from
the input field. Parameters are those of
Fig.~(\ref{fig:2ndary_bifudiag}). The cycle goes through the point
$F=0$. } \label{fig:limitcycle}
\end{figure}

The most intriguing transition is that one separating region $II$ from region
$III$. If one looks at the projection of the limit cycle of $F$ in the complex
plane ($f_s$, $f_c$), a crossing of the origin appears for a specific 
$\mu$-value where $F$ is instantaneously equal to 0. 
This can be seen in Fig.~\ref{fig:limitcycle}.

Since the $F$ dynamics is a limit cycle, it can be considered 
as a closed, oriented curve in the complex plane ($f_s$, $f_c$). 
It is thus possible to assign to it a winding number $n$ around 
the origin of the complex plane. In turns, this allows to define 
the average frequency of $F$ as $2\pi n/T$ where $T$ is the period of the cycle.
Therefore, the regimes in region $II$ and region $III$ are characterized by 
a zero and non zero average rotation, respectively.
One can consider intituively that the mechanism responsible for the red 
shift of the backward field with respect to the input field is at some
point responsible for shifting the forward field with respect to 
the frequency of the backward one. 

\begin{figure}[htp]
\centering
\includegraphics[width=0.95\linewidth, clip=true]{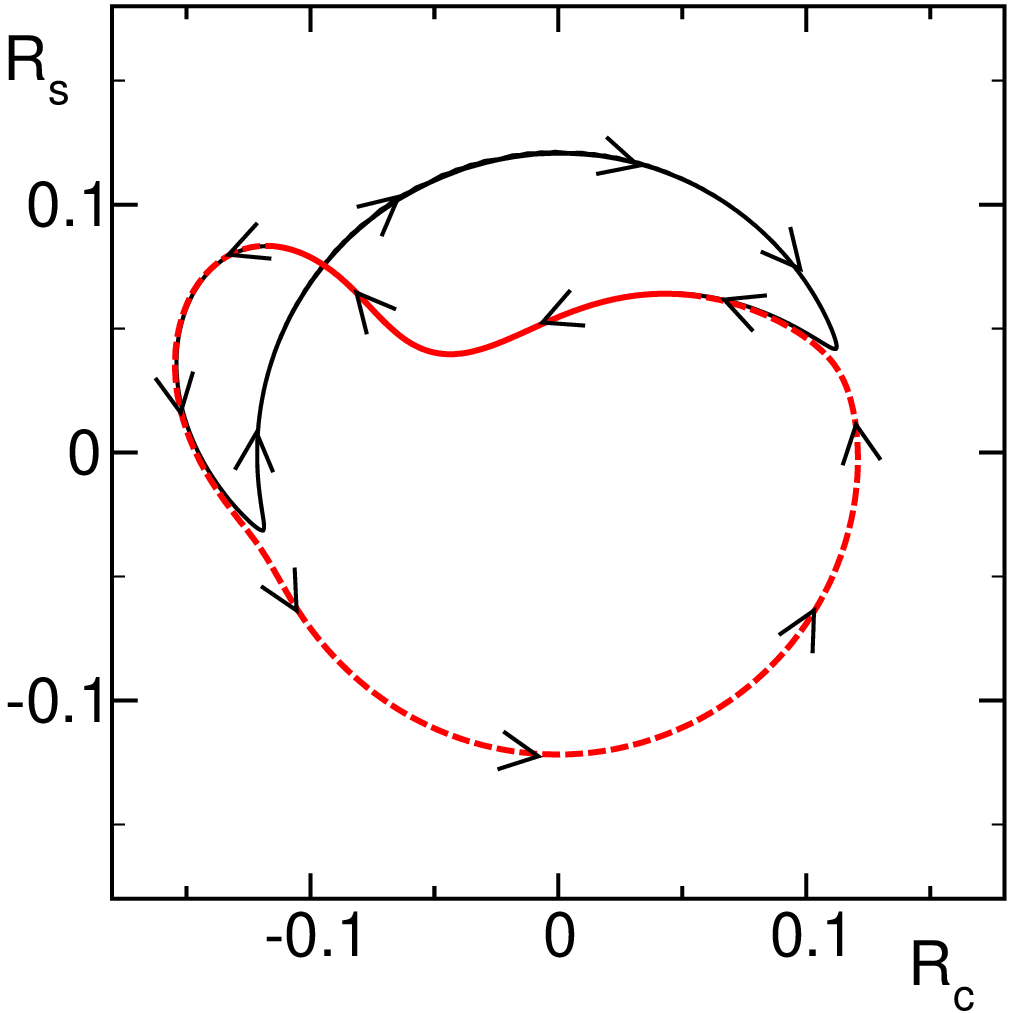}
\caption{ (color online) A projection of two limit cycles just below and above
the critical point, $\mu_c=2.752$ where the forward field unlocks
from the input field. The other parameter values are those of
Fig.~\ref{fig:cycle}. The arrows indicate the direction of the
motion: below (above) the transition, in the (un)locked regime,
the upper (lower) semicircle is followed. } \label{fig:rcscycle}
\end{figure}

\begin{figure}[htp]
\centering
\includegraphics[width=0.95\linewidth, clip=true]{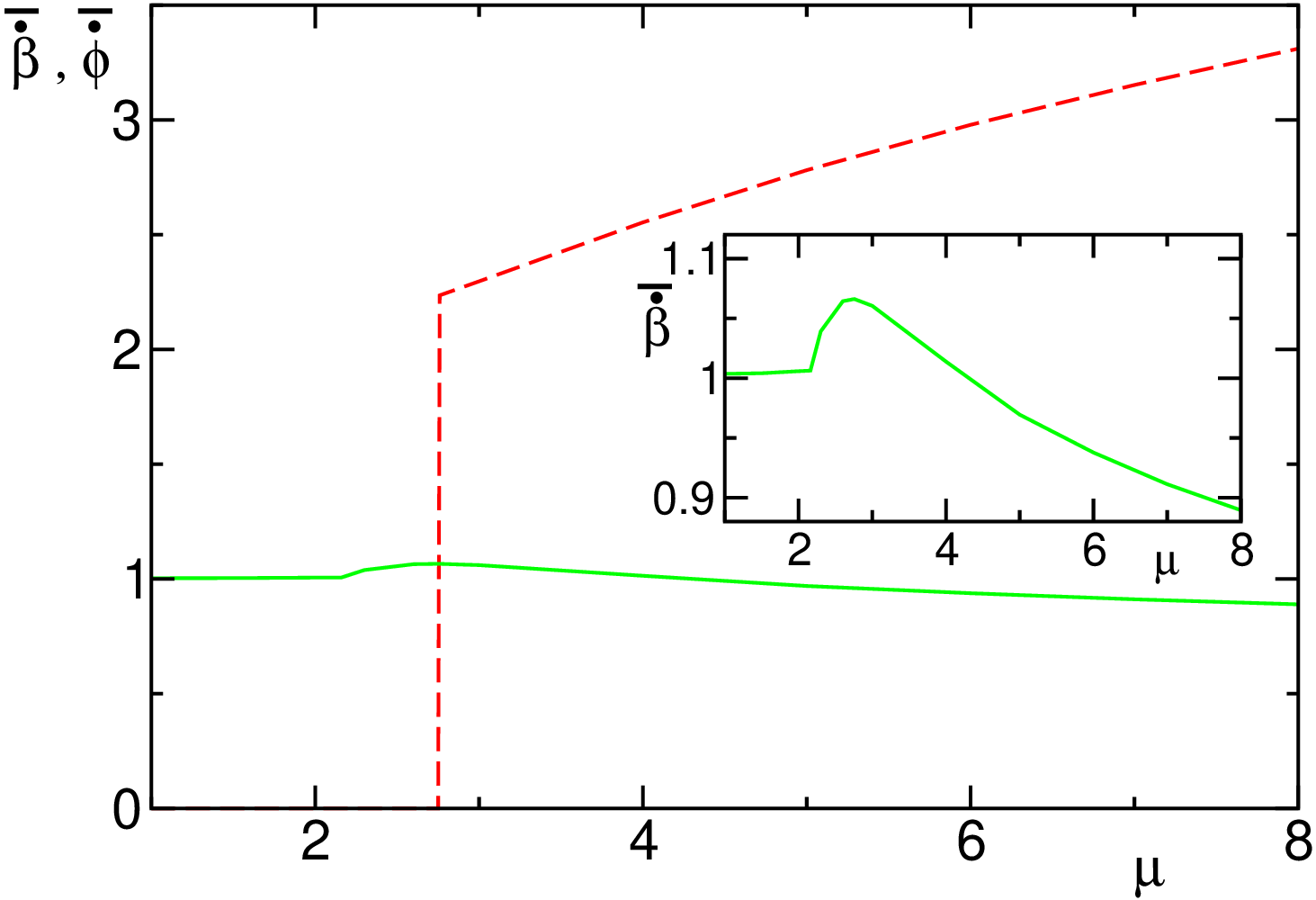}
\caption{ (color online) Frequency of the forward (dashed line) and backward
(solid line) fields, referred to that of the input field, as a
function of $\mu$. Parameters are the same as in Fig.~\ref{fig:2ndary_bifudiag}.
An enlargement of the $\overline{\dot \beta}$ behaviour is plotted in the inset.
} \label{fig:detun}
\end{figure}

A less trivial scenario is found in the ($R_c$, $R_s$) plane (see
Fig.~\ref{fig:rcscycle}): only a fraction of the limit cycle remains
unchanged when passing from below to above the transition point. The remaining
part is made of two complementary halves of a circumference, so that across
the transition, the whole limit cycle abruptly encircles the origin, thus
signaling  the onset of an order-parameter rotation.
In order to further clarify the transition, let us look at the evolution
of the potential tilt $\dot \phi -\dot \beta$.
\begin{figure}[htp]
\centering
\includegraphics[width=0.95\linewidth, clip=true]{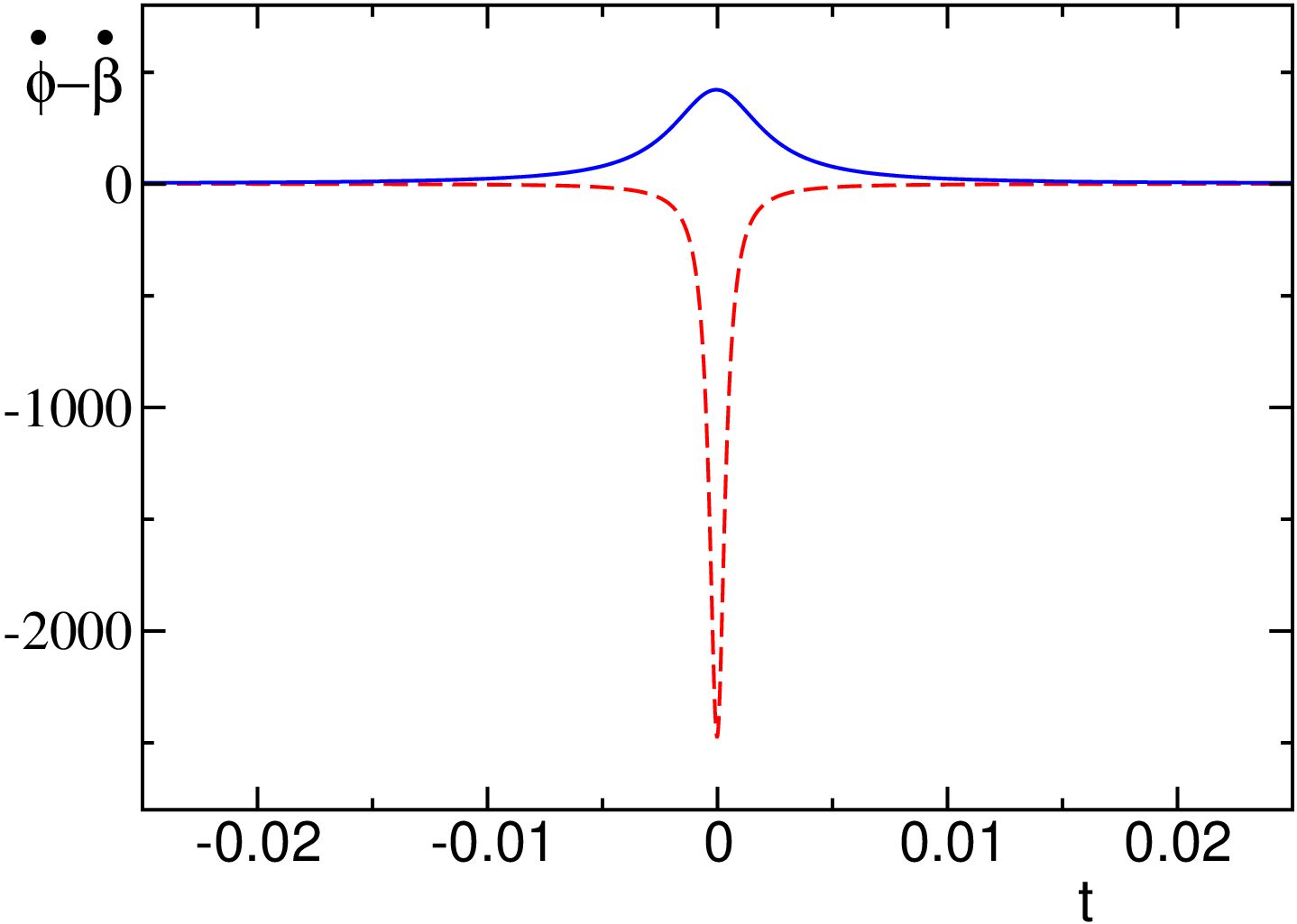}
\caption{ (color online) The behavior of the instantaneous slope of the
potential, just below ($\mu =2.75$ dashed line) and above ($\mu =
2.76$ the transition point). The different amplitude of the two
curves is due to a different distance from the critical point.
Parameters are those of Fig.~(\ref{fig:2ndary_bifudiag}).}
\label{fig:instome}
\end{figure}
In Fig.~\ref{fig:instome} we see that in the vicinity of the
singular point, where the field amplitude $f$ is close to zero
(for the sake of simplicity, we have set the origin of the time
axis such that the minimal distance from the origin occurs at
$t=0$), $\dot \phi - \dot \beta$ becomes very large, but the sign
of this quantity is different above and below the transition,
because the origin is encircled only above the transition. In the
laboratory frame, however, the average velocity of the atoms
exhibits a smooth change across the transition; in fact, the discontinuous
variation of the instantaneous frequency is
compensated by a discontinuous variation of the flux -- cf.
Eq.~(\ref{velocita}). Finally, in Fig.~\ref{fig:detun} we show the
dependency of the average frequency of both the forward
($\overline {\dot \phi}$), and the backward ($\overline {\dot \beta}$)
field, as a function of $\mu$.
There, we can see that both frequencies have the same sign and
that the red-shift of the forward field is larger than the one of
the backward field. Both features can be understood by means of the
following heuristic argument. We indeed expect that the same mechanism 
that is responsible for the red shift of the backward field with respect 
to the input field should, at some point (when the backward field intensity
is large enough) be responsible for red-shifting the forward field with respect 
to the backward one. This is precisely what we see.

\section{A minimal model} \label{sec_simp}

In this section we go back to the general case $\Delta \ne 0$ and show how
the model can be reduced to a set of five ordinary differential equations that is
still able to reproduce the relevant phenomenology discussed in the previous
section. We start by assuming that the parameter $C$ is large and
introduce the smallness parameter $\varepsilon \equiv C^{-1/3}$. As a
next step, we perform the following change of variables, that has
been suggested by numerical simulations,
\begin{eqnarray}
t &=& \varepsilon \tau, \nonumber \\
f &=& \varepsilon u \qquad b = \varepsilon v .
\end{eqnarray}
Moreover, we express the probability density $Q(\theta,t)$ as the sum of a
homogeneous component and a sinusoidal perturbation, namely,
\begin{equation}
Q(\theta,t) = \frac{1}{2\pi} + \varepsilon^2 S(\theta,\tau)
\end{equation}
and correspondingly define
\begin{equation}
R_s = \varepsilon^2 r_s \qquad  R_c = \varepsilon^2 r_c.
\end{equation}
Accordingly, the equations for the field variables write,
\begin{eqnarray}
\frac{d u}{d \tau} &=& \cos \phi - \varepsilon u - v r_s, \nonumber\\
\frac{d v}{d \tau} &=& -\varepsilon v + u r_s, \nonumber \\
\frac{d \phi}{d \tau} &=& -\frac{\sin \phi}{u} -\frac{v r_c}{u} -
\varepsilon \Delta, \label{eq:field2}\\
\frac{d \beta}{d \tau} &=& -\frac{u r_c}{v} -\varepsilon \Delta. \nonumber
\end{eqnarray}
while the Fokker-Planck equation writes,
\begin{eqnarray}
\partial_{\tau} S &=& - \frac{\mu \varepsilon}{2\pi} u v \cos \theta -
\partial_\theta \left[ \mathcal{D}\omega +
{\mu \varepsilon^3uv} \sin \theta\right] S \nonumber \\
 &&+ \sigma \varepsilon \partial^2_\theta S.
\label{eq:FP-new1}
\end{eqnarray}
where we have introduced
\begin{equation}
\mathcal{D} \omega = \frac{d \phi}{d \tau} - \frac{d \beta}{d
\tau} = r_c \left( \frac{u}{v}  -\frac{v}{u} \right) -\frac{\sin
\phi}{u}
\end{equation}
both to keep the notations as compact as possible and to remind that
$\beta$ is not a relevant variable.

So far, no approximation has been introduced, and the above two
sets of equations are equivalent to the initial formulation.
However, we can recognize the existence of small terms when
$\varepsilon$ is small. In particular, it is tempting to neglect
all terms which are proportional to some (positive) power of $\varepsilon$,
but this limit is singular. In fact, the resulting model is
dissipationless (notice also that all physical parameters would
disappear). On the one hand, the diffusion term in the
Fokker-Planck equation disappears as well as the position
dependent force, so that any initial condition for the
distribution $S(\theta)$ remains invariant in time, what is not
physical. On the other hand, the field dynamics is conservative as
well (the two loss terms vanish). Since, finally, as we see
below, there are conserved quantities, it is obvious that any
arbitrarily small dissipation is going to qualitatively modify the
asymptotic behavior and we, accordingly, cannot drastically set
the $\varepsilon$ terms equal to zero. Nevertheless, we are
entitled to neglect the cubic term, what is less crude an
hypothesis. The resulting simplified Fokker-Planck equation can be
solved exactly, assuming that $S(\theta)$ reduces to its first
Fourier mode. It is convenient to express the amplitude of such
mode directly referring to the two components of the order
parameter,
\begin{equation}
S(\theta) = \frac{1}{2 \pi}[ (r_c +i r_s) e^{-i \theta} +c.c. ]
\end{equation}
We indeed obtain (from now on, we again use a dot to mean the derivative
with respect to the time variable $\tau$)
\begin{eqnarray}
\dot r_c &=& -\frac{\mu \varepsilon}{2}uv - \mathcal{D}\omega r_s
- \sigma
\varepsilon r_c \nonumber \\
\dot r_s &=& \mathcal{D} \omega r_c -\sigma \varepsilon r_s
\label{eq:FP-reduce}
\end{eqnarray}
which complement the first three equations in Eqs.~(\ref{eq:field2})

If we now pass to phase and amplitude
\begin{equation}
r_c = r \cos \delta  \qquad r_s = r \sin \delta
\end{equation}
the entire set of equations writes
\begin{eqnarray}
\dot{u} &=& \cos \phi  -\varepsilon u - v r \sin \delta , \nonumber \\
\dot{v} &=& -\varepsilon v + u r \sin \delta \nonumber\\
\dot \phi &=& -\frac{rv}{u} \cos \delta - \frac{\sin \phi}{u} -\varepsilon
\Delta, \label{eq:FP-reduce-all} \\
\dot \delta &=& \mathcal{D}\omega +\frac{\mu \varepsilon}{2}
\frac{uv}{r} \sin \delta , \nonumber \\
\dot{r} &=& - \sigma \varepsilon r -\frac{\mu \varepsilon}{2} u v
\cos \delta \nonumber
\end{eqnarray}
while
\begin{equation}
\mathcal{D}\omega = r\cos\delta \left( \frac{u}{v} -\frac{v}{u}
\right) -\frac{\sin \phi}{u}
\end{equation}

Let us now discuss an analogy between our asymptotic limit and the small gain
approximation, or the Uniform Field Limit (UFL), that are widely used in laser
physics. Our approach implicitly assumes a weak action of the fields onto the
atomic sample,  i.e. a small dipolar coupling. Therefore, the Bragg grating
imprinted onto the atomic density can be considered as a small perturbation of
a homogeneous sample. On the other hand, one has to assume that the retroaction
of the atoms onto the cavity is sufficiently strong for the small density
modulation to influence  the fields, i.e. a large cooperation parameter $C$.
This approximation can be compared for instance with the small gain
approximation, where one assumes that the material gain is simply proportional
to its population inversion. This amounts to neglecting nonlinear saturation
processes such as power broadening for a two-level atoms or carrier heating
for a semiconductor material. Altogether, the asymptotic limit of this section
is analogous to the UFL which amounts to considering a weak single pass
gain within the cavity and, simultaneously, small optical losses,
in such a way that a finite net amplification can eventually occur.

\subsection{Numerical and theoretical analysis}

\subsubsection {The primary transition with non-zero detuning}

\begin{figure}[htp]
\centering
\includegraphics[width=0.95\linewidth, clip=true]{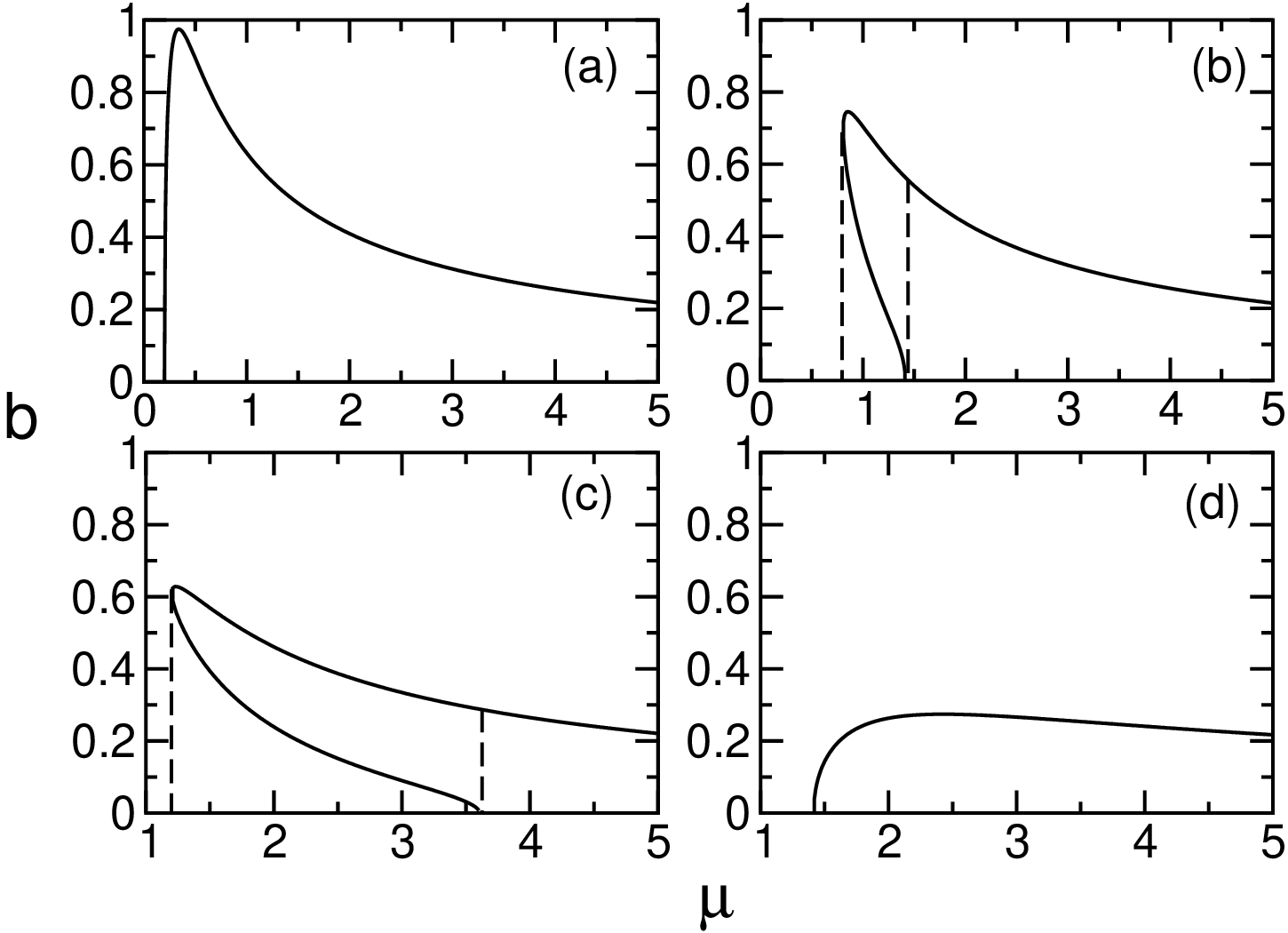}
\caption{Bifurcation diagram of the steady states of
Eqs.~(\ref{eq:FP-reduce-all}) as a function of $\mu$, for $C=20$ and $\sigma=1$.
Panels (a),(b),(c), and (d) correspond to $\Delta=0$, 2, 3, and -2, respectively.
The dashed lines denote the border of the bistability regions whenever there is
one.}
\label{fig:simp_bista}
\end{figure}

Analytic expressions  for the steady states of
Eqs.~(\ref{eq:FP-reduce-all}) are derived in the appendix. The
resulting bifurcation diagrams corresponding to $\Delta=0$, 2, 3,
and -2, respectively, are displayed in Fig.~\ref{fig:simp_bista}, 
where, for the
sake of clarity, we keep using the same notations as in the
previous section. In particular, we see that for positive and
large enough detuning there exist two branches (besides the
trivial one $b=0$). This means that the primary bifurcation
becomes subcritical, signaling the appearance of a  bistability region. 
Still
from the analytic discussion presented in the appendix, it turns
out that the primary threshold is
\begin{equation}
\label {r_results}
\mu_{th} = \frac{1+\Delta^{2}}{C}\left[
\Delta (\sigma - 1) + (\sigma + 1)\sqrt{\Delta^2  + 4\sigma }\right],\\
\end{equation}
It is important to notice that this expression holds true also for the
original model, since in the vicinity of the transition, the behaviour
of $Q(\theta)$ is by definition dominated by the first Fourier harmonic.

As shown in the appendix, it is possible to derive an analytic expression for
the saddle-node bifurcation, which turns out to be,
\begin{equation}
\label {r_results_fold}
\mu_{sn}  = \frac{2}{C}
\frac{\left(2\Delta\sigma-\Delta^{3}\right)\sqrt{\Delta^{2}+
4\sigma}+\Delta^{4}+2\sigma^{2}}{\sqrt{\Delta^{2}+4\sigma}-\Delta}.
\end{equation}
Notice that this equation makes sense only when both solutions of the
biquadratic equation (\ref{eq:biqua}) are positive. This can happen only for
$\Delta$ larger than a critical value $\Delta_c$ that can be determined by
setting $\mu_{sn}=\mu_{th}$,
\begin{eqnarray}
\Delta_{c} & = & \frac{1}{\sqrt{\sigma_{c}+1}}.
\end{eqnarray}
Thus, one can conclude that, when $\Delta$ is negative, no bistability
can occur, while it is allowed for a sufficiently large positive detuning.

\begin{figure}[htp]
\centering
\includegraphics[width=0.80\linewidth, clip=true]{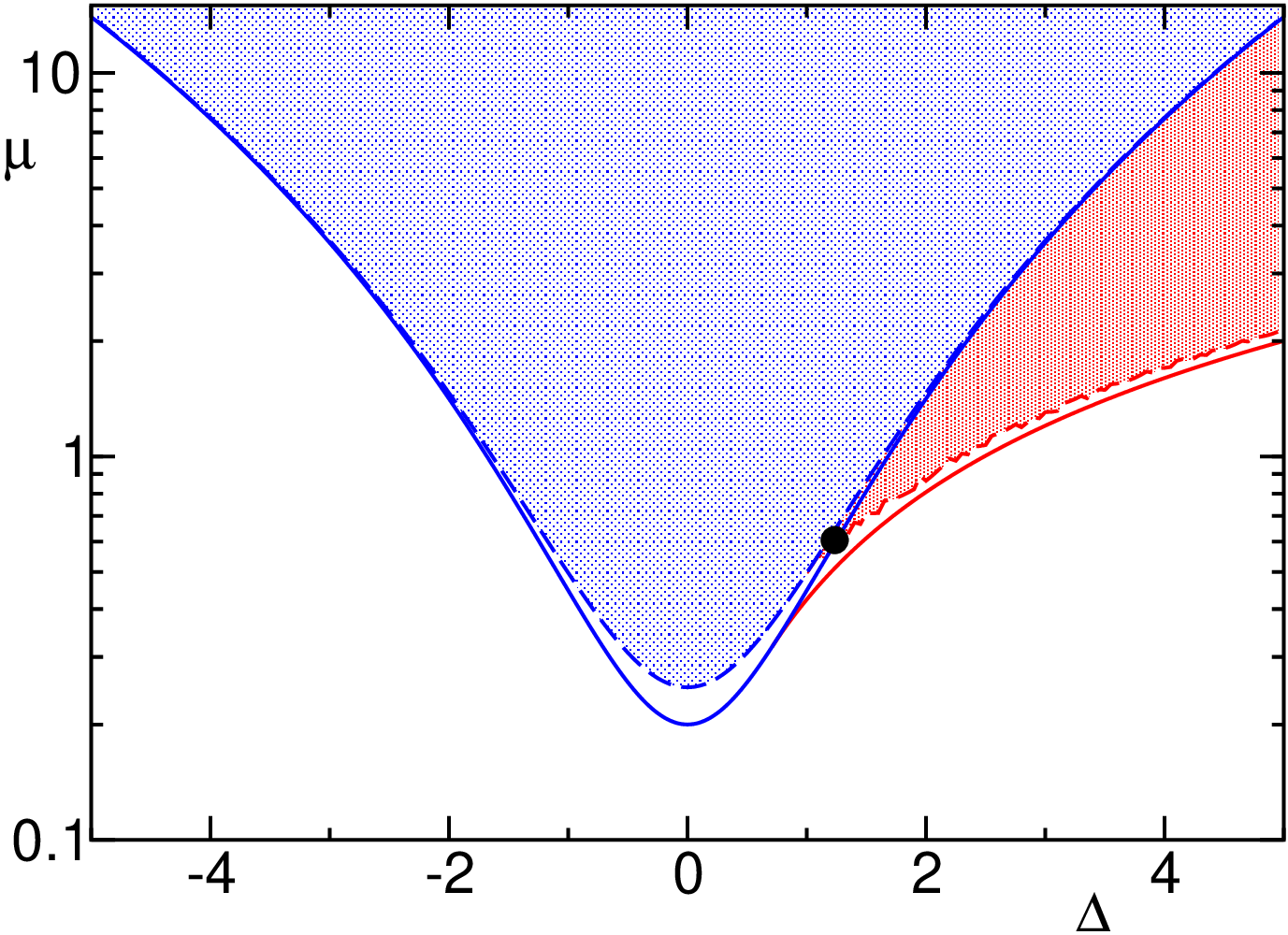}
\caption{ (color online) Spinodal decomposition of the steady
states. Dashed lines refer to the original full model (for $\sigma=1$ and
$C=20$); solid lines refer to the simplified model (\ref{eq:FP-reduce-all}).
The two shaded regions correspond to mono and bistable regimes in the full
model. The circle denotes the tricritical point $\Delta_c$.}
\label{fig:spino_simp_full}
\end{figure}

Finally, in the small temperature limit, the minimum threshold is achieved 
for a detuning
\begin{eqnarray}
\Delta_{m} & = & 2\frac{(1-\sigma)\sqrt{\sigma}}{1+9\sigma+8\sigma^{2}}.
\end{eqnarray}
Fig.~\ref{fig:spino_simp_full} shows the spinodal decomposition of
the solutions curve in the $(\Delta,\mu)$ plane of both the original and
the simplified model (see dashed and solid lines, respectively) for
$\sigma=1$ and $C=20$. One can see that there is a reasonable agreement
even though the corresponding $\varepsilon$-value is not too small ($\sim 0.36$).
The two shaded regions correspond to the mono- and the bistable regime, 
respectively.  The full circle marks the tricritical point, where the 
bistable area appears in the original model.

\subsubsection {The secondary transition}

\begin{figure}[htp]
\centering
\includegraphics[width=0.95\linewidth, clip=true]{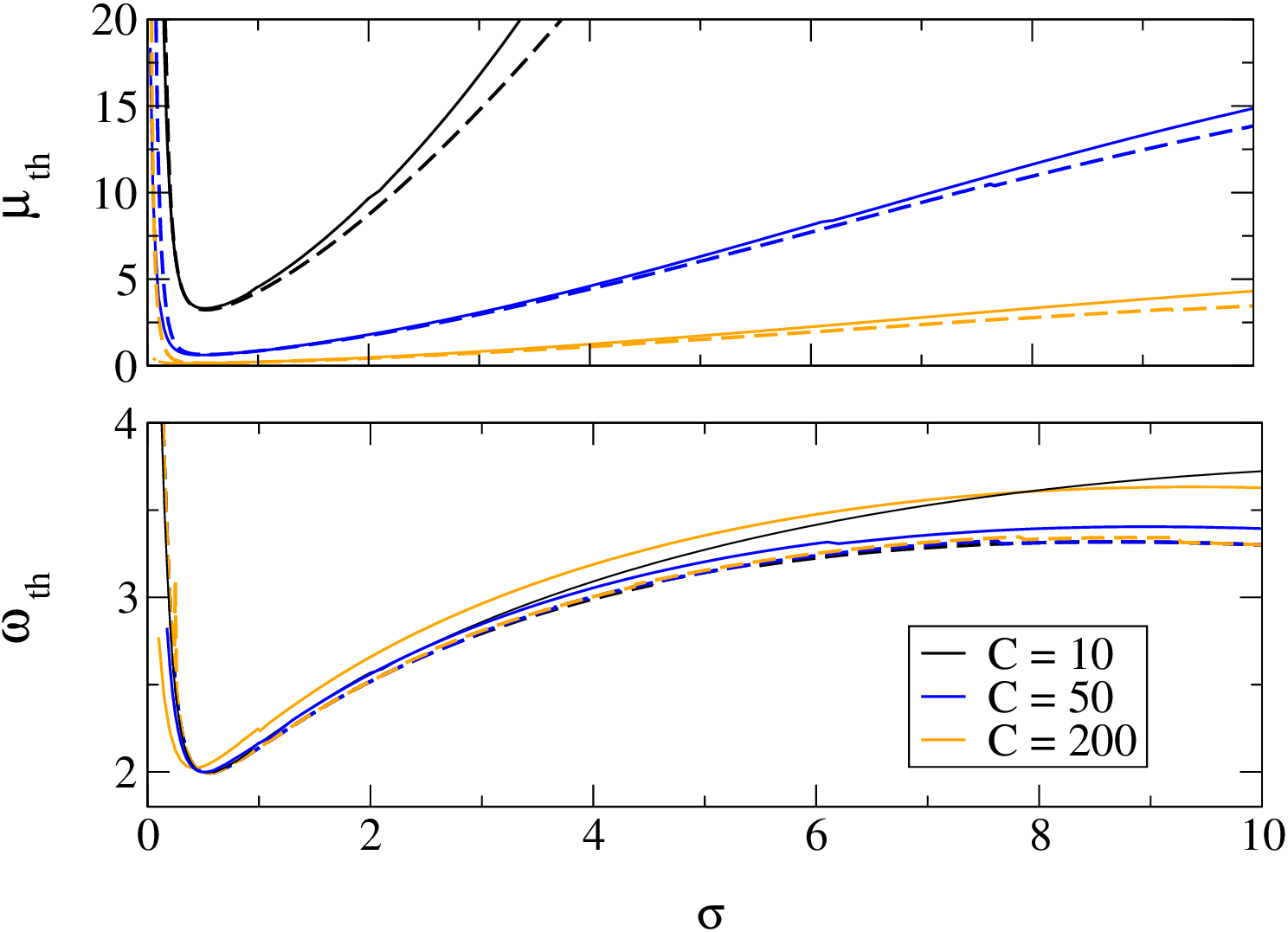}
\caption{(color online) Locus of the secondary instability as a function $\mu$
and $\sigma$, for different values of $C$. Predictions of the
simplified model, Eqs.~(\ref{eq:FP-reduce-all}) with dashed lines,
are compared with those of the exact model,
Eqs.~(\ref{ampf})-(\ref{fopl}), with solid lines. Parameters are
those of Fig. \ref{fig:2ndary_hopf}}
\label{fig:2ndary_hopf_simple}
\end{figure}

The almost quantitative agreement between the full and the simplified model 
is not
solely restricted to steady states. At larger input intensities, 
the simplified model exhibits a scenario that is much reminiscent 
of that observed in the original model: a secondary instability is first 
detected, that is followed by the unlocking phenomenon and, finally, 
by a sequence of period doubling bifurcations towards a chaotic regime.
This indicates that the degrees of freedom that are responsible for the onset
of macroscopic order are the very same ones leading to the 
self-pulsating instability.
In order to confirm whether the simplified model is really built on 
the relevant physical variables, we have investigated whether the locus 
of the secondary transition exhibits a similar dependence on the 
control parameters. We proceded along the same lines described 
in Sec.~\ref{subsec_second}.
Our main results are summarized in Fig.~\ref{fig:2ndary_hopf_simple}. 
There, one can see that the characteristics of the full model described in 
Fig.~\ref{fig:2ndary_hopf} are preserved, like e.g., inhibition 
of the transition for small values of either the temperature or the 
cooperation parameter $C$. By comparing the full and dashed lines 
in Fig.~\ref{fig:2ndary_hopf_simple}, one can also notice that an 
almost quantitative agreement with the full model is obtained whenever 
the value of $\mu_{th}$ is not too large.  Otherwise, the approximation, that 
consists in truncatiing the Fourier expansion of the probability distribution 
after the first mode, breaks. The overall quantitative agreement is reasonably 
good for values down to $C=50$, regardless of the value of the temperature 
$\sigma$.

\subsection{The $C \to \infty$ limit}

Since the highest degree of dynamical complexity amounts to a sequence
of period-doubling bifurcations, one might wonder whether it is possible
to further reduce the dimensionality of the model from five to three degrees
of freedom. In order to clarify this question we now consider the
limit $\varepsilon \rightarrow 0$ and, for the sake of simplicity, we restrict
the analysis to the resonant case $\Delta=0$. It is, unfortunately, necessary
to perform a further change of variables; namely we introduce
\begin{equation}
u_c = u \cos \phi  \qquad ; \qquad  u_s = u \sin \phi
\end{equation}
and
\begin{equation}
v_c = v \cos(\delta -\phi) \qquad ; \qquad v_s = v \sin(\delta -\phi)
\end{equation}
The resulting equations read
\begin{eqnarray}
\label{eq:rescal3}
\dot{r} &=& - \varepsilon \sigma r -\frac{\mu \varepsilon}{2}(v_c u_c
-v_s u_s) \nonumber \\
\dot{u_c} &=& 1 - r v_s  -\varepsilon u_c \nonumber \\
\dot{u_s} &=& - r v_c  -\varepsilon u_s \\
\dot{v_c} &=& ru_s -\varepsilon v_c -\frac{\mu \varepsilon}{2} \frac{v_s}{r}
(v_s u_c +v_c u_s) \nonumber \\
\dot{v_s} &=& r u_c -\varepsilon v_s +\frac{\mu \varepsilon}{2} \frac{v_c}{r}
(v_s u_c +v_c u_s) \nonumber
\end{eqnarray}

The great advantage of this representation is that in the $\varepsilon \to 0$
limit, it factorizes into three independent and partially degenerate blocks
\begin{eqnarray}
\label{eq:eps1}
\dot{r} &=& 0 \nonumber \\
\ddot{u}_c &=& - r^2 u_c \\
\ddot{u}_s &=& - r^2 u_s \nonumber
\end{eqnarray}
characterized by three constants of motion,
\begin{eqnarray}
\mathcal{C}_1 &=& r \nonumber \\
\mathcal{C}_2^2 &=& (v_s-1/r)^2+ u_c^2
\label{eq:constmo}\\
\mathcal{C}_3^2 &=& v_c^2 + u_s^2 \nonumber
\end{eqnarray}
Accordingly, a general solution writes as
\begin{eqnarray}
\label{eq:fastslow}
v_s &=& \frac{1}{\mathcal{C}_1} + \mathcal{C}_2 \sin ( \mathcal{C}_1t+\eta_1 ) \nonumber \\
u_c &=& \mathcal{C}_2 \cos ( \mathcal{C}_1 t+\eta_1 ) \nonumber \\
v_c &=& \mathcal{C}_3 \sin ( \mathcal{C}_1 t+\eta_2)  \\
u_s &=& \mathcal{C}_3 \cos ( \mathcal{C}_1 t+\eta_2) \nonumber
\end{eqnarray}
from which it is clear that two other constants enter the game, namely the phases
$\eta_1$, and $\eta_2$. However, one phase can be removed by shifting
the origin of the time axis, namely by introducing the phase difference
$\eta=\eta_2-\eta_1$.
Thus, we see that altogether, it should be possible at least to remove
one out of the 5 variables. However, rather than pursuing this goal,
we prefer to limit our discussion to the problem of determining the
value of all the relevant constants. In order to do that, it is necessary
to reintroduce a finite smallness parameter $\varepsilon$ and thereby
determining the time derivative of the various ``constants". By denoting with
a prime the derivative with respect to $\varepsilon\tau$, we find
\begin{eqnarray}
\label{eq:newmod}
\mathcal{C}'_1 &=& -\sigma \mathcal{C}_1 - \frac{ \mu}{2} \mathcal{C}_2
\mathcal{C}_3 \sin \eta \nonumber \\
\mathcal{C}'_2 &=& - \mathcal{C}_2 + \frac{ \mu \mathcal{C}_3}{16 \mathcal{C}_1^3}
\left[(\mathcal{C}_2^2-\mathcal{C}_3^2)\mathcal{C}_1^2-4\right] \sin \eta\\
\mathcal{C}'_3 &=& - \mathcal{C}_3 - \frac{ \mu \mathcal{C}_2}{16 \mathcal{C}_1^3}
\left[(\mathcal{C}_2^2-\mathcal{C}_3^2)\mathcal{C}_1^2+4\right] \sin \eta \nonumber \\
\eta' &=& -\frac{ \mu}{16
\mathcal{C}_1^3}\left[(\mathcal{C}_2^2+\mathcal{C}_3^2)\mathcal{C}_1^2+4\right]
   \left( \frac{\mathcal{C}_3}{\mathcal{C}_2} +
   \frac{\mathcal{C}_2}{\mathcal{C}_3} \right) \cos \eta \nonumber
\end{eqnarray}
These equations admit a pair of symmetric fixed points which correspond to a
periodic dynamics in the original variables,
\begin{equation}
\eta = \pm \frac{\pi}{2} \quad \mathcal{C}_1 = \pm \left( \frac{\mu}{4} \right)^{1/3} \quad
\mathcal{C}_2^2=\mathcal{C}_3^2 = \frac{\sigma}{2} \left( \frac{4}{\mu} \right)^{2/3}
\end{equation}
Therefore, the most robust dynamics which persists in the $C \to \infty$ limit,
is the periodic one, arising after the Hopf bifurcation.

\section{Conclusion} \label{sec_end}

By recasting a known CARL model as a self-consistent equation for the probability
distribution, we have been able to discuss analogies and differences
with the Kuramoto model for synchronization in ensemble of globally coupled
rotators. In fact, although the primary transition, giving rise to the
spontaneous formation of a density grating, resembles the onset of a
macroscopic synchronized state in the Kuramoto model, there are important
differences. In particular, the global coupling affects the frequency of the
oscillators (here, the velocity of the atoms), determining the tilting of
the effective washboard potential. Another difference concerns the existence,
in the CARL context, of an ``absolute" reference frame, the only one where 
the equations are time-independent. As a result of these differences, 
we find a subtlety in
the macroscopic behaviour: the average velocity of the grating does not
coincide with the average velocity of the single atoms.
Such a feature is reminiscent of the collective behaviour discussed in
\cite{PR07}, where it is shown that nonlinear all-to-all interactions may
lead to the onset of a peculiar periodic macroscopic phase. That phase
is (in the absence of external noise) both characterized by a microscopic 
quasi-periodic motion and an average frequency of the single oscillators 
that differs from the period of the macroscopic motion. 
However, the correspondence between this behaviour and the collective atomic 
motion arising beyond the primary threshold is perhaps incomplete, since in
the CARL context, the periodic global dynamics reduces to a fixed point 
in the ``absolute" reference frame. A more promising candidate to establish 
a full analogy is the periodic motion arising beyond the unlocking transition,
although the presence of microscopic stochastic fluctuations makes it 
difficult to formulate a convincing final statement. In fact, on the one
hand, quasiperiodicity can be recognized as such only in deterministic 
systems, and, on the other hand, we are aware of at least another mechanism 
leading to periodic oscillations in globally coupled noisy rotators 
(see, e.g. \cite{BCM87}). Unfortunately, in the context of cold atoms, 
we cannot consider directly the zero-noise limit, as it corresponds to a 
qualitatively different regime, namely full synchronization. 
Therefore, until an objective criterion of distinguishing possibly different
classes of periodic collective motions will be introduced, the problem
will remain open.

Finally, we wish to recall that the rich phenomenology extensively
discussed in this paper is experimentally accessible, since we have
everywhere (with the only exception of the zero-temperature limit) considered
parameter values that are compatible with the experiment discussed in
\cite{KCZC03}. The only important constraint comes from the need 
to stabilize and control a priori of the frequency of the input field, 
without the help of any feedback coming from the output of the 
cavity itself as done in Ref.~\cite{KCZC03}.

\section*{Appendix}
In this appendix we determine the steady states of Eqs.~(\ref{eq:FP-reduce-all}),
by setting all the derivatives equal to zero. From the second and the last of
them, one obtains
\begin{eqnarray}
\sin \delta &=& \frac{\varepsilon v}{u r} \nonumber \\
\cos \delta &=& -\frac{2\sigma r}{\mu u v}
\label{eq:delta}
\end{eqnarray}
By now subtracting the fourth from the third equation in
the set (\ref{eq:FP-reduce-all}) and thereby eliminating $\sin \delta$ and
$\cos \delta$ with the help of Eq.~(\ref{eq:delta}), we obtain a biquadratic
equation in $s=v/r$. Such an equation has always one and only one positive,
and thus physically acceptable, solution,
\begin{equation}
\frac{\varepsilon \mu}{2}s^2 = -\frac{\Delta}{2} +
\frac{\Delta}{2}\sqrt{\Delta^2 + 4 \sigma} .
\label{eq:s2}
\end{equation}
By now squaring and summing the two expressions for $\sin \delta$ and
$\cos \delta$ in (\ref{eq:delta}) we find that the intensity of the forward
field is
\begin{equation}
u^2 = \varepsilon^2 s^2 + \frac{4\sigma^2}{\mu^2}\frac{1}{s^2}
\label{eq:u}
\end{equation}
The last step consists in solving the first and third equation in
(\ref{eq:FP-reduce-all}) for $\cos \phi$  and $\sin \phi$, respectively,
squaring and summing. As a result, we obtain a biquadratic equation
for $r$,
\begin{equation}
 c_4 r^{4}+ c_2 r^2+ c_0 = 0
\label{eq:biqua}
\end{equation}
where
\begin{eqnarray}
c_4 & = &\varepsilon^2 s^4 + \frac{4 \sigma^2}{\mu^2} \nonumber \\
c_2 & = & 2u^2\varepsilon \left( \varepsilon s^2 - \frac{2\Delta \sigma}{\mu}\right)
\nonumber \\
c_0 &=& u^2 \left( \varepsilon^2 u^2 (1+\Delta^2) -1 \right) \nonumber
\end{eqnarray}
The bifurcation point of the homogeneous state $r=0$ is found by setting
$c_0=0$. With the help of Eqs.~(\ref{eq:u},\ref{eq:s2}), this condition
transforms into Eq.~(\ref{r_results}), displayed in Sec.~\ref{sec_simp}.

Moreover, the above biquadratic equation may have two distinct nontrivial
solutions. The critical point where a pair of solutions is
created (the saddle-node bifurcation) can be determined by imposing
\begin{equation}
c_2^{2}-4c_4 c_0  = 0
\end{equation}
With the help of Eqs.~(\ref{eq:u},\ref{eq:s2}), this condition
transforms into Eq.~(\ref{r_results_fold}). Notice that this condition
makes sense only when the two resulting solutions are both larger than
zero, i.e. when $c_2/c_4<0$.
Finally, from the way the solutions have been derived, one can observe
a curious property: the two branches are characterized by the same amplitude
of the forward field!

\acknowledgments

We wish to thank G.L. Lippi for useful discussions. One of us (AP) acknowledges financial
support from the Max Planck Institute for Complex Systems (Dresden) where this work was 
initiated.

\end{document}